\documentclass[a4paper,11pt]{article}
\pdfoutput=1 % if your are submitting a pdflatex (i.e. if you have
\usepackage{jcappub} % for details on the use of the package, please
\usepackage[T1]{fontenc} % if needed
\graphicspath{{images/}}
\usepackage{subcaption}
\usepackage{xspace}
\usepackage{multirow}
\usepackage{adjustbox}

\newcommand{\lya}{Lyman-$\alpha$\xspace}
\newcommand{\mpc}{~$h^{-1}$~Mpc\xspace}
\newcommand{\rp}{$r_\parallel$\xspace}
\newcommand{\rt}{$r_\perp$\xspace}
\newcommand{\drp}{$\Delta r_\parallel$\xspace}
\newcommand{\mpch}[1]{$#1\,h^{-1}\,\mathrm{Mpc}$}

\usepackage{float}

\title{Impact of Systematic Redshift Errors on the Cross-correlation of the \lya Forest with Quasars at Small Scales Using DESI Early Data}

\author[1,a]{Abby~Bault,\note[a]{Corresponding author.}}
\affiliation[1]{Department of Physics and Astronomy, University of California, Irvine, 92697, USA}

\author[1]{David~Kirkby,}

\author[2]{Julien~Guy,}
\affiliation[2]{Lawrence Berkeley National Laboratory, 1 Cyclotron Road, Berkeley, CA 94720, USA}

\author[3]{Allyson~Brodzeller,}
\affiliation[3]{Department of Physics and Astronomy, The University of Utah, 115 South 1400 East, Salt Lake City, UT 84112, USA}

\author[2]{J.~Aguilar,}

\author[4]{S.~Ahlen,}
\affiliation[4]{Physics Dept., Boston University, 590 Commonwealth Avenue, Boston, MA 02215, USA}

\author[2]{S.~Bailey,}

\author[5]{D.~Brooks,}
\affiliation[5]{Department of Physics \& Astronomy, University College London, Gower Street, London, WC1E 6BT, UK}

\author[6]{L.~Cabayol-Garcia,}
\affiliation[6]{Institut de F\'{i}sica d’Altes Energies (IFAE), The Barcelona Institute of Science and Technology, Campus UAB, 08193 Bellaterra Barcelona, Spain}

\author[6]{J.~Chaves-Montero,}

\author[2]{T.~Claybaugh,}

\author[7,8,9]{A.~Cuceu,}
\affiliation[7]{Center for Cosmology and AstroParticle Physics, The Ohio State University, 191 West Woodruff Avenue, Columbus, OH 43210, USA}
\affiliation[8]{Department of Physics, The Ohio State University, 191 West Woodruff Avenue, Columbus, OH 43210, USA}
\affiliation[9]{The Ohio State University, Columbus, 43210 OH, USA}

\author[3]{K.~Dawson,}

\author[10]{R.~de la Cruz,}
\affiliation[10]{Departamento de F\'{i}sica, Universidad de Guanajuato - DCI, C.P. 37150, Leon, Guanajuato, M\'{e}xico}

\author[11]{A.~de la Macorra,}
\affiliation[11]{Instituto de F\'{\i}sica, Universidad Nacional Aut\'{o}noma de M\'{e}xico,  Cd. de M\'{e}xico  C.P. 04510,  M\'{e}xico}

\author[12]{A.~Dey,}
\affiliation[12]{NSF's NOIRLab, 950 N. Cherry Ave., Tucson, AZ 85719, USA}

\author[5]{P.~Doel,}

\author[3,9]{S.~Filbert,}

\author[6]{A.~Font-Ribera,}

\author[13,14]{J.~E.~Forero-Romero,}
\affiliation[13]{Departamento de F\'isica, Universidad de los Andes, Cra. 1 No. 18A-10, Edificio Ip, CP 111711, Bogot\'a, Colombia}
\affiliation[14]{Observatorio Astron\'omico, Universidad de los Andes, Cra. 1 No. 18A-10, Edificio H, CP 111711 Bogot\'a, Colombia}

\author[15,16,17]{E.~Gaztañaga,}
\affiliation[15]{Institut d'Estudis Espacials de Catalunya (IEEC), 08034 Barcelona, Spain}
\affiliation[16]{Institute of Cosmology \& Gravitation, University of Portsmouth, Dennis Sciama Building, Portsmouth, PO1 3FX, UK}
\affiliation[17]{Institute of Space Sciences, ICE-CSIC, Campus UAB, Carrer de Can Magrans s/n, 08913 Bellaterra, Barcelona, Spain}

\author[2]{S.~Gontcho A Gontcho,}

\author[6]{C.~Gordon,}

\author[10]{H.~K.~Herrera-Alcantar,}

\author[7,8,9]{K.~Honscheid,}

\author[18]{V.~Ir\v{s}i\v{c},}
\affiliation[18]{Kavli Institute for Cosmology, University of Cambridge, Madingley Road, Cambridge CB3 0HA, UK}

\author[7,8,9,19]{N.~G.~Kara{\c c}ayl{\i},}
\affiliation[19]{Department of Astronomy, The Ohio State University, 4055 McPherson Laboratory, 140 W 18th Avenue, Columbus, OH 43210, USA}

\author[20]{R.~Kehoe,}
\affiliation[20]{Department of Physics, Southern Methodist University, 3215 Daniel Avenue, Dallas, TX 75275, USA}

\author[2]{T.~Kisner,}

\author[2]{A.~Kremin,}

\author[2]{A.~Lambert,}

\author[2]{M.~Landriau,}

\author[21]{L.~Le~Guillou,}
\affiliation[21]{Sorbonne Universit\'{e}, CNRS/IN2P3, Laboratoire de Physique Nucl\'{e}aire et de Hautes Energies (LPNHE), FR-75005 Paris, France}

\author[2]{M.~E.~Levi,}

\author[6,22]{M.~Manera,}
\affiliation[22]{Departament de F\'{i}sica, Serra H\'{u}nter, Universitat Aut\`{o}noma de Barcelona, 08193 Bellaterra (Barcelona), Spain}

\author[7,9,19]{P.~Martini,}

\author[12]{A.~Meisner,}

\author[6,23]{R.~Miquel,}
\affiliation[23]{Instituci\'{o} Catalana de Recerca i Estudis Avan\c{c}ats, Passeig de Llu\'{\i}s Companys, 23, 08010 Barcelona, Spain}

\author[24]{P.~Montero-Camacho,}
\affiliation[24]{Department of Astronomy, Tsinghua University, 30 Shuangqing Road, Haidian District, Beijing, China, 100190}

\author[25]{J.~Moustakas,}
\affiliation[23]{Department of Physics and Astronomy, Siena College, 515 Loudon Road, Loudonville, NY 12211, USA}

\author[11]{A.~Muñoz-Gutiérrez,}

\author[26]{J.~Nie,}
\affiliation[26]{National Astronomical Observatories, Chinese Academy of Sciences, A20 Datun Rd., Chaoyang District, Beijing, 100012, P.R. China}

\author[10,27]{G.~Niz,}
\affiliation[27]{Instituto Avanzado de Cosmolog\'{\i}a A.~C., San Marcos 11 - Atenas 202. Magdalena Contreras, 10720. Ciudad de M\'{e}xico, M\'{e}xico}

\author[2,28]{N.~Palanque-Delabrouille,}
\affiliation[28]{IRFU, CEA, Universit\'{e} Paris-Saclay, F-91191 Gif-sur-Yvette, France}

\author[29,30,31]{W.~J.~Percival,}
\affiliation[29]{Department of Physics and Astronomy, University of Waterloo, 200 University Ave W, Waterloo, ON N2L 3G1, Canada}
\affiliation[30]{Perimeter Institute for Theoretical Physics, 31 Caroline St. North, Waterloo, ON N2L 2Y5, Canada}
\affiliation[31]{Waterloo Centre for Astrophysics, University of Waterloo, 200 University Ave W, Waterloo, ON N2L 3G1, Canada}

\author[32]{I.~P\'erez-R\`afols,}
\affiliation[32]{Departament de F\'isica, EEBE, Universitat Polit\`ecnica de Catalunya, c/Eduard Maristany 10, 08930 Barcelona, Spain}

\author[2,33,34]{C.~Poppett,}
\affiliation[33]{Space Sciences Laboratory, University of California, Berkeley, 7 Gauss Way, Berkeley, CA  94720, USA}
\affiliation[34]{University of California, Berkeley, 110 Sproul Hall \#5800 Berkeley, CA 94720, USA}

\author[35]{F.~Prada,}
\affiliation[35]{Instituto de Astrof\'{i}sica de Andaluc\'{i}a (CSIC), Glorieta de la Astronom\'{i}a, s/n, E-18008 Granada, Spain}

\author[6]{C.~Ram\'irez-P\'erez,}

\author[28,36]{C.~Ravoux,}
\affiliation[36]{Aix Marseille Univ, CNRS/IN2P3, CPPM, Marseille, France}

\author[37]{M.~Rezaie,}
\affiliation[37]{Department of Physics, Kansas State University, 116 Cardwell Hall, Manhattan, KS 66506, USA}

\author[38]{G.~Rossi,}
\affiliation[38]{Department of Physics and Astronomy, Sejong University, Seoul, 143-747, Korea}

\author[39]{E.~Sanchez,}
\affiliation[39]{CIEMAT, Avenida Complutense 40, E-28040 Madrid, Spain}

\author[40]{E.~F.~Schlafly,}
\affiliation[40]{Space Telescope Science Institute, 3700 San Martin Drive, Baltimore, MD 21218, USA}

\author[2]{D.~Schlegel,}

\author[41,42]{M.~Schubnell,}
\affiliation[41]{Department of Physics, University of Michigan, Ann Arbor, MI 48109, USA}
\affiliation[42]{University of Michigan, Ann Arbor, MI 48109, USA}

\author[2]{J.~Silber,}

\author[28]{T.~Tan,}

\author[42]{G.~Tarl\'{e},}

\author[43,44]{M.~Walther,}
\affiliation[43]{Excellence Cluster ORIGINS, Boltzmannstrasse 2, D-85748 Garching, Germany}
\affiliation[44]{University Observatory, Faculty of Physics, Ludwig-Maximilians-Universit\"{a}t, Scheinerstr. 1, 81677 M\"{u}nchen, Germany}

\author[12]{B.~A.~Weaver,}

\author[26]{Z.~Zhou}

\emailAdd{abault@uci.edu}

\abstract{The Dark Energy Spectroscopic Instrument (DESI) will measure millions of quasar spectra by the end of its 5 year survey. Quasar redshift errors impact the shape of the \lya forest correlation functions, which can affect cosmological analyses and therefore cosmological interpretations. Using data from the DESI Early Data Release and the first two months of the main survey, we measure the systematic redshift error from an offset in the cross-correlation of the \lya forest with quasars. We find evidence for a redshift dependent bias causing redshifts to be underestimated with increasing redshift, stemming from improper modeling of the \lya optical depth in the templates used for redshift estimation. New templates were derived for the DESI Year 1 quasar sample at $z > 1.6$ and we found the redshift dependent bias, \drp, increased from $-1.94 \pm 0.15$\mpc to $-0.08 \pm 0.04$\mpc ($-205 \pm 15~\text{km s}^{-1}$ to $-9.0 \pm 4.0~\text{km s}^{-1}$). These new templates will be used to provide redshifts for the DESI Year 1 quasar sample.}

\begin{document}

\maketitle
\flushbottom

\section{Introduction}
\label{sec:intro}
The Dark Energy Spectroscopic Instrument (DESI) \cite{desi2016a, desi2016b, desikp1} saw its first light in October 2019 and began its main survey in May of 2021. Prior to the main survey, DESI collected spectra in its Survey Validation (SV) \cite{desiSV} phase between December 2020 and May 2021, which make up the data for the Early Data Release (EDR) \cite{edr}. The EDR consists of approximately 1.7 million unique spectral objects among which 90,000 are quasars. Over the course of 5 years, DESI will collect spectra from 40 million extragalactic objects, with approximately 3 million of those spectra coming from quasars \cite{desiSV, edr}. 

Quasars, or quasi-stellar objects (QSO), offer access to the highest extragalactic redshifts that can be observed and measured by DESI. Indeed, they are among the most luminous sources in the universe and as such they have been used for many years to study the large-scale structure of the universe as well as probing the expansion of the universe using Baryon Acoustic Oscillations (BAO) \cite{ebossqso, ebossqso2, dmdb2020, boss2013, bossbaodr9, bossdr11}. The BAO scale, a standard ruler once normalized to the sound horizon $r_\text{d}$, allows for constraints on cosmological parameters \cite{cuceu2022, ebosscosmo}.

At high redshifts ($z > 2.0$), in order to study the large-scale matter distribution, DESI is not only utilizing quasar positions, but also the \lya forest in their spectra. The \lya forest is created when the light of a distant quasar passes through and is absorbed by intervening neutral Hydrogen gas in the intergalactic medium (IGM). The light from the quasar is redshifted as it travels and once it is redshifted to the wavelength of the \lya transition (1215.67 \AA), it will be absorbed by the neutral hydrogen in the IGM. This occurs anytime the light is redshift to the \lya transition, thus creating the ``forest''. Using both quasars and the \lya forest as tracers, it is possible to measure the BAO scale to high precision. This was already done by the eBOSS collaboration \cite{dmdb2020}. DESI recently also measured 3D correlations with quasars and the \lya forest using EDR data \cite{calum2023}, though the results on BAO were not reported. Beyond BAO, high-redshift quasar spectra are also used, for example to measure the 1D power spectrum (P1D) of the \lya forest \cite{corentin, naim}, or to study metals in the intergalactic medium \cite{naim2}.

In order to get the high precision needed for the cosmological measurements from BAO with the \lya forest-quasar cross-correlation, the redshifts that DESI measures for each quasar spectrum need to be extremely precise. While the redshift measurement of the majority of DESI galaxies, which rely on the presence of narrow emission lines and spectral features, is indeed very precise, this is not the case for quasars whose emission lines are essentially broad, and complex.  There are therefore many factors that can cause errors on redshift measurements. Quasars that have Broad Absorption Line (BAL) features tend to have larger errors on redshift since the BAL features tend to affect the absorption blueward of the emission line centers. The resulting emission line asymmetry can shift the redshift estimate \cite{garcia2023, bal1}. Additionally, quasar emission line profiles can vary from quasar to quasar, and can be shifted from the systemic redshift. This is especially true for high-ionization, broad emission lines like \texttt{CIV}, which can be blueshifted a few hundred km~s$^{-1}$ from low-ionization, broad emission lines like \texttt{MgII} which are often closer to the systemic redshift \cite{gaskell, shen2016}. Emission lines can also be broadened, and even asymmetric, for example \lya absorption in the IGM can cause an asymmetry in the line profile \cite{richards2002, gaskell}. 

After the spectra are collected and processed by the DESI spectroscopic pipeline \cite{guy}, they are classified and the redshifts are estimated by the redshift fitter code adopted by the DESI collaboration, Redrock \cite{redrock}. Redrock is both a redshift fitter and classifier, and fits to pre-computed Principal Component Analysis (PCA) templates to spectra to determine the most probable class of the object (galaxy, quasar, star) and the best-fit redshift. The quasar spectral templates used by Redrock for EDR were adopted from eBOSS and are described in \cite{bolton}. They were developed using a small number of spectra, and can not fully account for the variability of spectral features in quasar spectra. The performance of these templates on DESI EDR quasars is discussed in \cite{qsovi} and \cite{brodzeller2023}. 

It was reported in \cite{brodzeller2023} that there is a redshift-dependent bias in the eBOSS quasar templates likely stemming from the improper correction of the \lya forest optical depth which introduces suppression in the forest that increases with redshift. This bias is present in the quasar catalogs used in this work since the eBOSS templates were used to produced the EDR redshifts. In section \ref{sec:zdep}, we discuss mitigating this bias by accounting for the mean transmission of \lya in the spectral templates used for redshift estimation. 

We can utilize the \lya forest-quasar cross-correlation to constrain quasar redshift errors due to its asymmetry along the line of sight. When measuring the cross-correlation as a function of the parallel and perpendicular components of the line of sight separation (\rp and \rt, respectively), a positive offset from \rp~$= 0$, meaning the cross-correlation is shifted to a higher redshift than that of the quasar, indicates that quasar redshifts are being underestimated. As shown in \cite{youles2022}, quasar redshift errors smear the BAO peak in the radial direction and cause unphysical correlations for small separations in \rt that increase with increasing redshift errors. However, \cite{youles2022} also found that the position of the BAO peak is resilient to effects from quasar redshift errors.

In this work we measure and study the \lya forest-quasar cross-correlation, focusing on scales less than \mpch{80}. We begin in section~\ref{sec:data}, where we present and describe the sample of quasars and \lya forests used in this analysis. We describe the method used to measure and fit the correlations in section~\ref{sec:method}. In section~\ref{sec:mocks}, we discuss validating our method with mocks. We present the cross-correlation and baseline fit results in section~\ref{subsec:corrfit}, we explore any redshift evolution of the redshift error parameter in section \ref{subsec:zdep}, and the impact of Broad Absorption Line regions in section \ref{subsec:bal}. We address the redshift dependent bias on the quasar templates in section \ref{sec:zdep}. Finally, we conclude in section~\ref{sec:conclusions} with a summary of this work and suggestions for future studies.

\section{Data Sample}
\label{sec:data}
The data used in this paper consists of quasars from the DESI EDR \cite{edr} as well as data from the first two months of main survey observations (M2). The combination of EDR and M2 is hereafter referred to as EDR+M2. The data products from EDR are described in \cite{edr} and \cite{guy}, and the spectra are publicly available \footnote{\url{https://data.desi.lbl.gov/public/edr/}}. The spectra from M2 will be publicly available with the year 1 data release. We present the quasar sample in \ref{subsec:qso} and the \lya forest sample in \ref{subsec:lyaforest}.

\subsection{Quasar Sample}
\label{subsec:qso}

The quasar catalog was constructed following the logic in section 6.2 of \cite{edmond}, specifically figure 9. We briefly describe the method here.

The quasar catalog is produced using Redrock \cite{redrock} and the two afterburner algorithms: a broad Magnesium II (MgII) line finder and a machine-learning based classifier QuasarNET \cite{bb, qn, green}. Objects are first run through Redrock to determine both the class of the object (galaxy, quasar, star) and the best-fit redshift. Objects that are classified \texttt{galaxy} are then run through the MgII algorithm. If the MgII algorithm finds significant broad \texttt{MgII} emission, then their classification is changed to \texttt{QSO} and the redshift from Redrock remains unchanged. Any remaining objects are then run through QuasarNET. If QuasarNET classifies an object as \texttt{QSO}, Redrock is re-run for that object with the QuasarNET redshift as a prior. If neither of the three algorithms classify an object as \texttt{QSO}, then it is not included in the catalog. While the redshifts from QuasarNET are included in the catalog (as value-added content), the official redshift for each object comes from Redrock. 

\begin{figure*}[t!]
    \centering
    \begin{subfigure}[t]{0.49\textwidth}
        \centering
        \includegraphics[width = \textwidth]{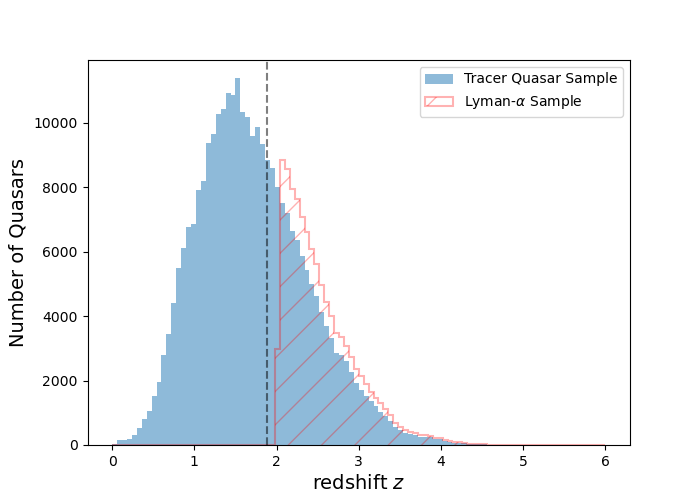}
        %\caption{}
        %\caption{Redshift distribution for the tracer quasar sample (without BALs) used in this paper. The red hatched histogram represents the \lya forest sample.  The dashed grey line represents the lower bound on the tracer quasar sample, $z=1.88$.}
        %\label{fig:zdist}
    \end{subfigure}
    \hfill
    \begin{subfigure}[t]{0.49\textwidth}
        \centering
        \includegraphics[width=\textwidth]{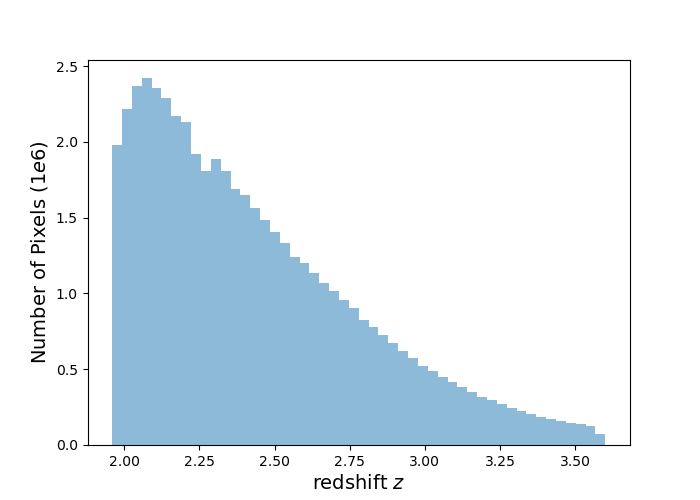}
        %\caption{Redshift distribution of \lya forest pixels output during the continuum fitting process for the baseline analysis. }
        %\caption{}
        %\label{fig:pixels}
    \end{subfigure}
    \caption{The left plot shows the redshift distribution for the tracer quasar sample (without BALs) used in this paper. The red hatched histogram represents the \lya forest sample. The dashed grey line represents the lower bound on the tracer quasar sample, $z = 1.88$. The right plot shows the redshift distribution of the \lya forest pixels output during the continuum fitting process for the baseline analysis.}
    \label{fig:zdistfootprint}
\end{figure*}

To finalize our catalog, we select only the objects that were targeted as quasars, using the classifications provided by the DESI target-selection pipeline \cite{meyers2023}. Some of these objects were observed multiple times during EDR and M2 and therefore have multiple entries, but we want to keep only one observation per object in the quasar catalog. To select the best observation for the duplicated quasars, we first remove those that have a redshift warning flags set (i.e, those with a bad fit). If there are still repeat observations of the same quasar at this point, we select the observation with the highest \texttt{TSNR2\_LYA} value. \texttt{TSNR2\_LYA}, which is defined in section 4.14 of \cite{guy}, is a measure related to the signal to noise ratio in the blue part of spectra for a given observation. We also remove any BAL quasars that were identified by the BALFinder \cite{mg2019} as having their absorption index (AI) > 0. 

After these cuts, a total of 290,506 quasars are present in the quasar catalog. However, only the 106,861 quasars with redshift z > 1.88 will be included in the cross-correlation discussed in section \ref{subsec:corrfunc}. We refer to these quasars as tracer quasars or the tracer quasar catalog. Figure~\ref{fig:zdistfootprint} shows the redshift distribution of the entire quasar sample with the dashed grey line representing this cutoff for the tracer quasar sample. The red hatched histogram shows the redshift distribution for the \lya sample discussed in the next section.

\subsection{\lya Forest Sample}
\label{subsec:lyaforest}

\begin{figure}
    \centering
    \includegraphics[width=\textwidth]{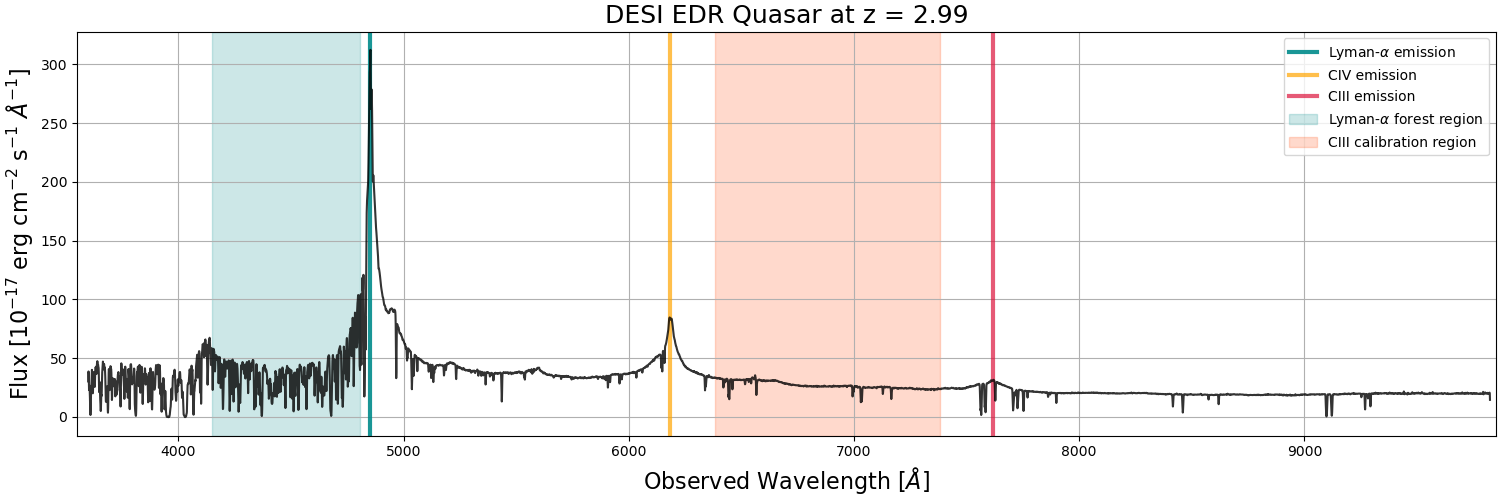}
    \caption{An example of a quasar spectrum from DESI EDR at redshift 2.99 and \texttt{TARGETID}~= 39633362754732929. The observed \lya emission line is shown in teal, \texttt{CIV} in orange and \texttt{CIII}] in red. The orange shaded region is the \texttt{CIII} calibration region, and the \lya forest region is shaded in teal, both are used in the continuum fitting process in section~\ref{subsec:contfit}. Redward of the \lya emission are some absorption lines, likely from various metals.}
    \label{fig:svqso}
\end{figure}

We use the same \lya forest sample as described in \cite{cesar}. We will briefly describe the sample here, but refer the reader to \cite{cesar} for a more detailed description. 

We define the \lya forest as the region between the Lyman-$\beta$ and \lya emission lines at 1025.7~\AA~and 1215.67~\AA, respectively. To reduce any contamination from the wings of those emission lines, we reduce the rest-frame wavelength range of all the \lya forests to 1040~\AA~< $\lambda$ < 1205~\AA. An example quasar spectrum highlighting the \lya forest is shown in figure \ref{fig:svqso}. Other features highlighted are the \texttt{CIV} and semi-forbidden \texttt{CIII} emission lines. The region between these lines shaded in orange is used during the calibration step of the continuum fitting process discussed in section \ref{subsec:contfit} and in \cite{cesar}. 

The \lya forest sample is created from the same quasar catalog described in section \ref{subsec:qso}, but the BAL quasars are not excluded. We set a limit on the observed wavelength range of the \lya forest between 3600~\AA~and 5772 \AA. The lower limit is set to the lowest wavelength measured by the blue DESI spectrographs \cite{guy} and corresponds to a minimum \lya pixel redshift of $z = 1.96$, meaning the lowest redshift a quasar can have is $z = 2.02$. The upper limit is set by the small number of high redshift quasars in EDR+M2. This limit corresponds to a maximum quasar redshift of $z$ = 4.29 (with a corresponding maximum \lya pixel redshift of $z= 3.75$). 

With these redshift cuts, there are 109,900 valid \lya forests from the quasar catalog. Prior to fitting the continuum (described in section \ref{subsec:contfit}), we first mask regions of the spectra that are contaminated. This contamination can come from sky lines, galactic absorption, BALs, and Damped Lyman Alpha (DLA) regions. We mask any DLA regions identified with the DLAFinder \cite{dla} and BAL regions identified with the BALFinder \cite{mg2019} of the quasar spectra. For the sky lines and galactic absorption, we mask four areas of the spectra: sky lines from [5570.5, 5586.7]~\AA; and the \texttt{Ca H}, and \texttt{K} lines from the Milky Way from [3967.3, 3971.0]~\AA~and [3933.0, 3935.8]~\AA, respectively. We also perform a calibration using the \texttt{CIII} between 1600 < $\lambda_\text{RF}$ [\AA] < 1850, which is shaded in orange in figure \ref{fig:svqso}. This calibration step follows the same process described in section \ref{subsec:contfit} and in \cite{cesar}. We require each forest to have a minimum length of 40 \AA. Forests that are too short (don't have enough pixels), have low signal (< 0), or have problems fitting the continuum (negative continuum) will be rejected during the calibration and continuum fitting process (section~\ref{subsec:contfit}). After these rejections, the final \lya sample has 88,511 forests. Figure~\ref{fig:zdistfootprint} shows the redshift distribution of \lya pixels in the sample.

\section{Method}
\label{sec:method}

We measure and model the cross-correlation of the \lya forest with quasars following the method in \cite{calum2023, dmdb2020} and for the continuum fitting process we follow the method in \cite{cesar}. We will briefly discuss the method for this analysis starting with the continuum fitting in section~\ref{subsec:contfit}, calculating the cross-correlation function in section~\ref{subsec:corrfunc}, the distortion and covariance matrices in section~\ref{subsec:distcov}, and the model in section~\ref{subsec:model}. We perform the calculations in sections \ref{subsec:contfit}-\ref{subsec:distcov} using \texttt{picca}\footnote{\url{https://github.com/igmhub/picca/}}, and we perform the modeling in section \ref{subsec:model} using \texttt{vega}.\footnote{\url{https://github.com/andreicuceu/vega}}

\subsection{Measuring the Flux Transmission Field}
\label{subsec:contfit}

We follow the method in \cite{cesar} to measure the flux transmission field for the \lya sample. We will briefly describe the method here but refer the reader to \cite{cesar} for a more detailed description.

%Prior to fitting the continuum for the quasars in the \lya sample, we first mask regions of the spectra that are contaminated. This contamination can come from sky lines, galactic absorption, BALs, and DLAs. For the sky lines and galactic absorption, we mask four areas of the spectra: sky lines from [5570.5, 5586.7]~\AA; and the Ca H, K, and Na lines from the Milky Way from [3967.3, 3971.0]~\AA, [3933.0, 3935.8]~\AA, and [5885.7, 5903.3]~\AA, respectively. As mentioned in section \ref{subsec:lyaforest}, we follow \cite{cesar} to mask the regions in the quasar spectra identified as a DLA by the DLAFinder \cite{dla} or BAL by the BALFinder \cite{mg2019}. We also perform a calibration using the the \texttt{CIII} between 1600 < $\lambda_\text{RF}$ [\AA] < 1850, which is shaded in orange in figure \ref{fig:svqso}. This calibration step follows the same process described in this section and in \cite{cesar}. 

In this work, we adopt the same 0.8 \AA~linear pixel size used by the DESI spectroscopic pipeline \cite{guy} and other \lya forest analyses using DESI EDR+M2 data \cite{calum2023, cesar}. Previous analyses in eBOSS used logarithmic pixels \cite{dmdb2020}. We follow the method described in \cite{cesar}. 

For each forest in each line of sight, $q$, we calculate the flux transmission field $\delta_\text{q}$ as a function of observed wavelength: 
\begin{equation}
    \label{eq:delta}
    \delta_\text{q}(\lambda) = \frac{f_\text{q}(\lambda)}{\overline{F}(\lambda)C_\text{q}(\lambda)} - 1.
\end{equation}
Here, $f_\text{q}(\lambda)$ is the observed flux, $C_\text{q}(\lambda)$ is the unabsorbed quasar continuum, and $\overline{F}(\lambda)$ is the mean transmitted flux at a specific wavelength. To measure each delta field, we estimate the mean expected flux, $\overline{F}(\lambda)C_\text{q}(\lambda)$, which is the product of the terms in the denominator in eq.~\ref{eq:delta}. We assume it can be expressed according to:
\begin{equation}
    \label{eq:deltaapprox}
    \overline{F}(\lambda)C_\text{q}(\lambda) = \overline{C}(\lambda_\text{RF})\left(a_\text{q} + b_\text{q}\frac{\Lambda - \Lambda_\text{min}}{\Lambda_\text{max} - \Lambda_\text{min}}\right),
\end{equation}
where  $\Lambda \equiv \log\lambda$, $\overline{C}(\lambda_\text{RF})$ is the estimate of the mean continuum for the considered quasar sample, and $a_\text{q}$, $b_\text{q}$ are per-quasar parameters.

The estimated variance of the flux $\sigma_\text{q}^2$ depends on the noise from the DESI spectroscopic pipeline and the intrinsic variance of the \lya forest ($\sigma_\text{LSS}^2$) and we model it as
\begin{equation}
    \label{eq:sigmaq}
    \frac{\sigma_\text{q}^2(\lambda)}{(\overline{F}C_\text{q}(\lambda))^2} = \eta(\lambda)\Tilde{\sigma}_\text{pip,q}^2 + \sigma_\text{LSS}^2,
\end{equation}
where $\Tilde{\sigma}_\text{pip,q} = \sigma_\text{pip,q}/\overline{F}C_\text{q}(\lambda)$ and $\sigma_\text{pip,q}$ is the spectroscopic pipeline estimate of the flux uncertainties. It is further modified by the correction, $\eta(\lambda)$, which accounts for inaccuracies in the spectroscopic pipeline noise estimation. In DESI $\eta(\lambda)$ ranges between 0.97 and 1.05 between 3600 \AA~and 5772 \AA, respectively, and is generally close to 1 due to first performing a calibration using the CIII region. The mean continuum $\overline{C}(\lambda_\text{RF})$, the function $\eta(\lambda)$, $\sigma_\text{LSS}(\lambda)$, and the quasar parameters, $a_\text{q}$ and $b_\text{q}$, are fit in an iterative process and stable values are obtained after about 5 iterations.

\subsection{The \lya Forest-Quasar Cross-Correlation}
\label{subsec:corrfunc}
In this work we study the \lya forest-quasar cross-correlation function. We require the use of the \lya auto-correlation function to constrain some of the fit parameters (discussed in section~\ref{sec:results}). We use the \lya forest auto-correlation results from \cite{calum2023} and, when necessary (in sections \ref{sec:mocks} and \ref{subsec:bal}), run the auto-correlation following the method in \cite{calum2023}. The results of the cross-correlation are presented in section~\ref{subsec:corrfit}. 

The separation between two tracers is determined both parallel and perpendicular to the line of sight, with $\vec{r}$ = (\rp, \rt). For two tracers \emph{i} and \emph{j} with redshift \emph{$z_\text{i}$} and \emph{$z_\text{j}$}, that are offset by an observed opening angle $\Delta\theta$, the comoving separation can be computed as
\begin{subequations}
    \begin{equation}
        \label{eq:rpar}
        r_\parallel = \left[D_\text{c}(z_\text{i}) - D_\text{c}(z_\text{j})\right]\cos{\left(\frac{\Delta\theta}{2}\right)},
    \end{equation}
    \begin{equation}
        \label{eq:rperp}
        r_\perp = \left[D_\text{m}(z_\text{i}) + D_\text{m}(z_\text{j})\right]\sin{\left(\frac{\Delta\theta}{2}\right)},
    \end{equation}
\end{subequations}
where $D_\text{c}(z) = c\int_0^z dz/H(z)$ is the comoving distance, $D_\text{m}$ is the angular comoving distance, and $H(z)$ is the Hubble function. To use the comoving distance we have to assume a fiducial cosmology and we adopt the flat $\Lambda$CDM cosmology of Planck Collaboration et al. (2018) \cite{planck}. With this cosmology we have $D_\text{c} = D_\text{m}$.
In the above equations, the two tracers are either two \lya pixels for the auto-correlation or a \lya pixel and a quasar for the cross-correlation. The redshift of a \lya pixel is calculated as $z_\text{i} = \lambda_\text{obs}/\lambda_\text{Ly$\alpha$} - 1$, where $\lambda_\text{Ly$\alpha$} = 1215.67$ \AA~and the quasar redshift is given from the tracer quasar catalog described in section \ref{subsec:qso}.

%\subsubsection{The lya-quasar cross-correlation}
%\label{subsubsec:xcorrfunc}

The cross-correlation of the \lya forest with quasars is defined as
\begin{equation}
    \xi(\vec{r}_\text{A}) = \langle \delta_\text{Q}(\vec{x})\delta(\vec{x} + \vec{r}_\text{A})\rangle \approx \langle \delta(\vec{x}_\text{Q} + \vec{r}_\text{A}) \rangle,
\end{equation}
where $\delta_\text{Q}(\vec{x})$ is the fluctuation of the number density in quasars, and $\vec{r} = (r_\parallel, r_\perp)$ (with $r^2 = r_\parallel^2 + r_\perp^2$) for bin A. If we assume that quasars are shot-noise dominated, then the cross-correlation is just the average \lya transmission around quasars at $\vec{x}_\text{Q}$ \cite{afr2012b}.

For the \lya-quasar cross-correlation we use the same estimator as previous studies \cite{dmdb2020, calum2023, afr2013}:
\begin{equation}
    \label{eq:xcorr}
    \xi_\text{A} = \frac{\sum_\text{(i,j)$\in$A}w_\text{i} w_\text{j} \delta_\text{i}}{\sum_\text{(i,j)$\in$A}w_\text{i} w_\text{j}},
\end{equation}
where $i$ indexes a \lya pixel and $j$ a quasar, and $\delta_\text{i}$ is the flux transmission for the \lya pixel defined in eq.~\ref{eq:delta}. The weights $w$ are defined in \cite{calum2023} and proportional to $(1+z)^{\gamma-1}$, where $\gamma = 1.44$ for quasars and 2.9 for a \lya pixel. The sums are over all quasar-pixel pairs with separations that are within the bounds of bin A, with the exception of \lya pixels from their own background quasar which are excluded. The bounds of bin A are defined in comoving coordinates and are determined based on the selected bin width. 

The correlation is binned along and across the line of sight, but the cross-correlation itself is not symmetric (i.e, you can't swap indices $i$ and $j$ for $\delta_i$ in eq.~\ref{eq:xcorr}). We define a positive separation along the line of sight when the quasar is in front of the \lya pixel ($z_{qso} < z_{Ly\alpha}$) and likewise define a negative separation along the line of sight when the quasar is behind the \lya pixel ($z_{qso} > z_{Ly\alpha}$). This asymmetry in the cross-correlation is what allows us to study systematic quasar redshift errors.

In this work we are studying small-scale effects on the cross-correlation, so we do not need to extend to large values of $r_\parallel$ and $r_\perp$, unlike the studies of BAO and the full-scale correlations in \cite{calum2023, dmdb2020}. We instead use a maximum separation of \mpch{80} in both \rp and \rt. For the cross-correlation, this corresponds to a minimum tracer quasar redshift of $z = 1.88$ given the minimum \lya pixel redshift of $z = 1.96$ given in section \ref{subsec:lyaforest}. 

Since our maximum separation in this study is \mpch{80}, for the cross-correlation we restrict \rp $\in [-80, 80]$\mpc and \rt $\in [0, 80]$\mpc. Previous measurements of 3D correlations to study BAO with eBOSS or DESI \cite{calum2023, dmdb2020} used a bin width of \mpch{4}. Here, the focus of this paper is to measure a small-scale effect, we use a bin width of \mpch{1}, along both transverse and longitudinal separations. This gives us 80 bins across the line of sight, 160 bins along the line of sight, and N = 80 $\times$ 160 = 12800 bins in total. We validate our choice to use a \mpch{1} bin width with mock data in section \ref{sec:mocks}.

\subsection{The Distortion and Covariance Matrices}
\label{subsec:distcov}
We follow the same method as in \cite{calum2023} to estimate the covariance matrix for the cross-correlation. The covariance matrix is estimated by splitting the sky into $s$ regions (according to \texttt{HEALPix} \cite{healpix} pixels) and then computing their weighted covariance. For our analysis we use 440 \texttt{HEALPix} pixels with \texttt{nside} = 16. Assuming that correlations between regions are negligible, the covariance is written as
\begin{equation}
    \label{eq:cov}
    C_\text{AB} = \frac{1}{W_\text{A}W_\text{B}}\sum_\text{s} W^\text{s}_\text{A} W^\text{s}_\text{B}[\xi^\text{s}_\text{A} \xi^\text{s}_\text{B} - \xi_\text{A}\xi_\text{B}],
\end{equation}
where A and B are two different bins of the correlation function $\xi$, $W=\sum_\text{s} W^\text{s}$, $W^\text{s}$ is the summed weight for one of the regions $s$, and $\xi_\text{A} = \sum_\text{s} W_\text{A}^\text{s} \xi_\text{A}^\text{s}/W_\text{A}$. 

The cross-covariance matrix has N = $12800 \times 12800 = 163$ million entries and the auto-covariance (from \cite{calum2023}) has N = $2500 \times 2500 = 6.25$ million entries. The covariance matrices are dominated by the diagonal elements of the matrix. The estimates of off-diagonal elements are noisy, so they are modeled as a function of the difference of separation along the line of sight ($\Delta r_{\parallel} = |r^\text{A}_{\parallel} - r^\text{B}_{\parallel}|$) and across the line of sight ($\Delta r_{\perp} = |r^\text{A}_{\perp} - r^\text{B}_{\perp}|$) and then smoothed \cite{dmdb2020}. This smoothing is necessary to obtain a positive-definite, or invertible, covariance matrix. Correlation coefficients that have the same $\Delta r_{\parallel}, \Delta r_{\perp}$ are averaged.  

Using the \lya forest when estimating the quasar continuum biases the estimates of the delta field in eq.~\ref{eq:delta} because the measured delta for a given pixel is a combination of the absorption signals located at other pixels. This produces a distortion in the correlation functions that occurs throughout, but mainly appear for small \rt. The biases arise because \begin{enumerate}
    \item fitting the quasar parameters $a_\text{q}$ and $b_\text{q}$ biases the mean $\delta_\text{q}$ toward 0, and 
    \item fitting the mean transmitted flux, $\overline{F}$, biases the mean $\delta$ at each observed wavelength, $\overline{\delta(\lambda)}$, toward 0.
\end{enumerate} 
These effects can be modeled by transformations to $\delta_\text{q}(\lambda)$. With the assumption that these transformations to $\delta_\text{q}$ are linear, the correlation function is distorted by a distortion matrix, $D_\text{AA'}$, such that 
\begin{equation}
    \label{eq:distmat}
    \xi^\text{distorted}_\text{A} = \sum_\text{A'} D_\text{AA'}\xi^\text{true}_\text{A'}.
\end{equation}
For the cross-correlation, the distortion matrix is calculated as
\begin{equation}
    \label{distmat_cross}
    D_\text{AA'}^\text{cross} = \frac{\sum_\text{(i,k)$\in$A} w_\text{i} \sum_\text{(j,k)$\in$A'} \eta_\text{ij}}{\sum_\text{(i,k)$\in$A} w_\text{i}},
\end{equation}
where $\eta$ is the projection due to the distortion that occurs during the continuum fitting process and is defined in equation 3.3 of \cite{calum2023}, and the weights $w$ are defined in eq. \ref{eq:xcorr}. 

When fitting the data, (described in section~\ref{subsec:model}), the physical model of the correlation function is multiplied by the distortion matrix.

\subsection{The Model of the \lya Forest-Quasar Cross-correlation}
\label{subsec:model}

This section presents the theoretical model for the cross-correlation. Though this work focuses on the cross-correlation, the auto-correlation is necessary to include during the fitting process as it helps break degeneracies between parameters. We follow the method from \cite{calum2023} for fitting the auto-correlation, and use the auto-correlations from \cite{calum2023} in the baseline analysis.

In this model, the expected measured (distorted) cross-correlation in the (\rp, \rt) bin $A$ is related to the theoretical (true) correlation by the distortion matrix as shown in eq.~\ref{eq:distmat}. The theoretical cross-correlation, $\xi^\text{qf,th}$, can be broken down into the sum of its components:
\begin{equation}
    \label{eq:model_components}
    \xi^\text{qf,th} = \xi^\text{Ly$\alpha\times$QSO} + \sum_\text{a} \xi^\text{QSO$\times$a} + \xi^\text{TP}.
\end{equation}
Here, the first term gives the contribution from the cross-correlation between the \lya forest and quasars, the second term gives the contribution from the cross-correlation between quasars and other absorbers, like metals and high column density systems (HCDs), in the \lya region, and the last term gives the contributions from the transverse proximity (TP) effect, which is the effect of radiation from the quasar on the nearby surrounding gas. We describe these terms in depth throughout this section and the parameters of this model are described in table~\ref{table:modelparams}.

%Power spectra and bias paramters

The \lya-QSO contribution to the correlation function is derived from the Fourier transform of the tracer biased power spectrum, $P(\textbf{k},z)$, which is written as 
\begin{equation}
    \label{eq:powerspec}
    P(\textbf{k},z) = b_\text{i}(z) b_\text{j}(z)(1 + \beta_\text{i} \mu_\text{k}^2)(1 + \beta_\text{j} \mu_\text{k}^2) P_\text{QL}(\textbf{k},z) F_\text{NL}(\textbf{k}) G(\textbf{k}),
\end{equation}
where $\textbf{k}$ is the wavenumber with modulus $k$ and $\mu_\text{k} = k_\parallel/k$. The bias and linear redshift-space distortion parameters, $b_\text{i,j}(z)$ and $\beta_\text{i,j}$, are for tracer $i$ or $j$. $G$ is a correction that accounts for the averaging of the correlation function binning and $F_\text{NL}$ accounts for the non-linear effects on small scales (large \textbf{k}). $P_\text{QL}$ is the quasi-linear matter power spectrum defined in eq.~\ref{eq:PQL}. For the auto-correlation, $i=j$ and the only tracer is \lya absorption. In this work, we are concerned with the cross-correlation, $i \neq j$ and the two tracers are \lya absorption and quasars.

%bias & RSD parameters
In eq.~\ref{eq:powerspec}, the bias $b$ and redshift-space distortion $\beta$ parameters for each tracer $i,j$ appear with the standard Kaiser factor: $b_\text{i,j}(z)(1+\beta_\text{i,j}\mu_\text{k}^2)$ \cite{kaiser}. The fit of the cross-correlation is only sensitive to the product of the quasar and \lya biases. The quasar bias $b_\text{q}$ is redshift dependent and is given by 
\begin{equation}
    \label{eq:bq_zdep}
    b_\text{q}(z) = b_\text{q}(z_\text{eff})\left(\frac{1+z}{1 + z_\text{eff}}\right)^{\gamma_\text{q}},
\end{equation}
where $z_\text{eff}$ is the effective redshift of the sample and $\gamma_\text{q}$ = 1.44 \cite{dmdb2019}. Following the analysis of \cite{calum2023} we allow $b_\text{q}$ to remain free in the fit. On the other hand, the quasar redshift-space distortion (RSD) parameter, $\beta_\text{q}$, is not a free parameter. It is related to the quasar bias $b_\text{q}$ by: 
\begin{equation}
    \label{eq:rsdqso}
    \beta_\text{q} = \frac{f}{b_\text{q}},
\end{equation}
where $f = 0.9703$ is the linear growth rate of structure in our fiducial cosmology. We then derive the value of $\beta_\text{q}$ from the result of $b_\text{q}$. 

We assume that the \lya bias parameter, $b_\text{Ly$\alpha$}$, is redshift dependent but use the approximation that the \lya RSD parameter, $\beta_\text{Ly$\alpha$}$, is redshift independent. Following \cite{calum2023}, $b_\text{Ly$\alpha$}$ is given by
\begin{equation}
    \label{eq:blya}
    b_\text{Ly$\alpha$}(z) = b_\text{Ly$\alpha$}(z_\text{eff})\left(\frac{1+z}{1+z_\text{eff}}\right)^{\gamma_\text{Ly$\alpha$}},
\end{equation}
with $\gamma_\text{Ly$\alpha$}$ = 2.9 \cite{mcdonald2006}. We allow $b_\text{Ly$\alpha$}$ and $\beta_\text{Ly$\alpha$}$ to remain free in the fit.

The \lya absorption contains contributions from both the inter-galactic medium (IGM) and high-column-density (HCD) systems. HCD absorbers, such as Damped Lyman Alpha, trace the underlying density field. When these HCD systems are correctly identified and the appropriate parts of the absorption region are properly masked and modeled, they will have no effect on the measured correlation functions. However, if these systems are left unidentified, they will affect the measured correlation function as a broadening/smearing in the radial direction. Following the method in \cite{afr2012}, this effect introduces a $k_\parallel$ dependence to the effective bias and can be modeled as
\begin{equation}
    \label{eq:effbias}
    \begin{split}
        b'_{\text{Ly}\alpha} &= b_{\text{Ly}\alpha} + b_{\text{HCD}}F_{\text{HCD}}(k_\parallel),\\
        b'_{\text{Ly}\alpha}\beta'_{\text{Ly}\alpha} &= b_{\text{Ly}\alpha}\beta_{\text{Ly}\alpha} + b_{\text{HCD}}\beta_{\text{HCD}}F_{\text{HCD}}(k_\parallel),
    \end{split}
\end{equation}
where $(b_{\text{Ly}\alpha}, \beta_{\text{Ly}\alpha})$ and $(b_\text{HCD}, \beta_\text{HCD})$ are the bias and redshift space distortion parameters associated with the IGM and HCD systems. Following \cite{dmdb2020}, we use $F_\text{HCD} = \exp(-L_\text{HCD}k_\parallel)$, which is approximated from the results in \cite{rogers2018}. $L_\text{HCD}$ is the length scale for unmasked HCDs, and is degenerate with other parameters. For this work, we fix $L_\text{HCD} =$ \mpch{10} and allow $b_\text{HCD}$ to be free in the fit. We fix $\beta_\text{HCD} = 0.5$.

While this work focuses on comoving separations smaller than the BAO scale, for consistency with other analyses \cite{calum2023,desikp6}, we model the quasi-linear matter power spectrum, $P_\text{QL}(\textbf{k},z)$, as the sum of a smooth and peak components with an empirical anisotropic damping applied to the BAO peak component:
\begin{equation}
    \label{eq:PQL}
    P_\text{QL}(\textbf{k},z) = P_\text{sm}(k,z) + A_\text{peak}\exp{\left[-\frac{k^2_\parallel\Sigma^2_\parallel + k^2_\perp\Sigma^2_\perp}{2}\right]}P_\text{peak}(k,z).
\end{equation}
$P_\text{sm}$ in eq.~\ref{eq:PQL} is derived from the linear power spectrum $P_\text{L}(k,z)$ via the side-band technique described in \cite{kirkby2013}. The redshift dependent linear power spectrum, $P_\text{L}$, is derived from CAMB \cite{camb}, and we then define $P_\text{peak} = P_\text{L} - P_\text{sm}$. The BAO peak is affected by non-linear broadening \cite{eisenstein2007} and is corrected for by introducing the parameters $(\Sigma_\parallel, \Sigma_\perp)$. $\Sigma_\parallel$ is related to $\Sigma_\perp$ and the growth rate such that
\begin{equation}
    \label{eq:sigma_f}
    \frac{\Sigma_\parallel}{\Sigma_\perp} = 1 + f,
\end{equation}
where $f = 0.9703$ is the linear growth rate of structure in our fiducial cosmology. Given that we do not fit the BAO peak in this work, we fix ($A_\text{peak},~\Sigma_\parallel,~\Sigma_\perp$) = (1, \mpch{6.37}, \mpch{3.24}) \cite{calum2023}.

$F_\text{NL}$ in equation~\ref{eq:powerspec} corrects for additional effects at small scales. In the auto-correlation these are thermal broadening, peculiar velocities, and nonlinear growth structure, parameterized according to equation 3.6 in \cite{arinyo2015}. For the cross-correlation the most important small-scale correction is due to quasar velocities and the precision of quasar redshift measurements. Following \cite{calum2023, dmdb2017}, we model this using the Lorentz-damping form:
\begin{equation}
    \label{eq:FNL}
    F_\text{NL}^\text{cross}(k_\parallel) = \frac{1}{\sqrt{1+(k_\parallel\sigma_\text{v})^2}},
\end{equation}
where $\sigma_\text{v}$ is a free parameter that describes the precision of quasar redshift measurements. Alternative models for $F_\text{NL}$ use a Gaussian form. 

The last term in eq.~\ref{eq:powerspec}, $G(\textbf{k})$, accounts for the effects of the binning of the correlation function on the $(r_\parallel, r_\perp)$ separation grid. Assuming the distribution in each bin is homogeneous, and following the method used in \cite{bautista2017},
\begin{equation}
    \label{eq:Gk}
    G(\textbf{k}) = \text{sinc}\left(\frac{R_\parallel k_\parallel}{2}\right)\text{sinc}\left(\frac{R_\perp k_\perp}{2}\right),
\end{equation}
where $R_\parallel$ and $R_\perp$ are the scales of the smoothing, i.e, are the radial and transverse widths of the bins, respectively. In this work we focus on bin widths of \mpch{1} for both  $R_\parallel$ and $R_\perp$, but validate the choice between \mpch{1} and \mpch{4} in section \ref{sec:mocks}.

%\subsubsection{Absorption by metals}
The next component that contributes to the model cross-correlation in eq.~\ref{eq:model_components} is the sum over the non-\lya absorbers, or metals. The power spectrum for the cross-correlation with metals has the same form as that used for the \lya-quasar cross-correlation, though it is simplified by neglecting the effect of HCDs. Because there is little absorption by metals, it is further simplified by not separating the smooth and peak components. It is however, more complicated due to the fact that the (\rp, \rt) bins in the correlation correspond to an observed $(\Delta\theta, \Delta\lambda)$ calculated assuming that the absorption is due to the \lya transition. 

This causes a shift in the model correlation function, and the contribution of each absorber is maximized in the (\rp, \rt) bin that corresponds to vanishing physical separation. For \lya absorption this corresponds to (\rp, \rt) = (0,0). For the other absorbers, this maximum occurs at \rt = 0 and $r_\parallel \approx \frac{c}{H(z)}(1+z)(\lambda_\text{m} - \lambda_{\text{Ly}\alpha})/\overline{\lambda}$, where $\overline{\lambda}$ is the mean value between the $\text{Ly}\alpha$ and metal absorption \cite{calum2023}. The values of the \rp separations for the metal absorbers considered in this work are given in table~\ref{table:metals}.

\begin{table}[]
    \centering
    \begin{tabular}{cccc}
    
    Transition & $\lambda_\text{m}$~[\AA] & $\lambda_\text{m}/\lambda_{\text{Ly}\alpha}$ & \lya-\text{m} $r_\parallel$~[\mpc] \\ \hline
    SiII(1260) & 1260.4 & 1.036 & +104 \\
    SiIII(1207) & 1206.5 & 0.992 & -22 \\
    SiII(1193) & 1193.3 & 0.981 & -55 \\
    SiII(1190) & 1190.4 & 0.979 & -62 \\

    \end{tabular}
    \caption{Metal transitions seen in the IGM and that are also present in the \lya forest-quasar cross correlation for $r_\parallel \in [-80, 80]$~\mpc. The second column gives the rest-frame wavelength of the transition in Angstroms. The third column is the ratio between the wavelengths of the metal transition and the \lya transition (1215.67 \AA). The last column gives the error in comoving distance when a metal absorption is incorrectly assumed to be \lya absorption at $z = 0$.}
    \label{table:metals} 
\end{table}

The shifted-model correlation function is calculated with respect to the unshifted-model correlation function, $\xi^{mn}$, for each absorber pair $(m, n)$ by introducing a metal matrix $M_\text{AB}$ as in \cite{blomqvist2018}:
\begin{equation}
    \label{eq:metalcorr}
    \xi^\text{mn}_\text{A} \longrightarrow \sum_\text{B} M_\text{AB}\xi^\text{mn}(r_\parallel(B),r_\perp(B)),
\end{equation}
where $M_\text{AB}$, the metal matrix, is defined as
\begin{equation}
    \label{eq:metalmat}
    M_\text{AB} = \frac{1}{W_\text{A}}\sum_\text{(m,n)$\in$A,(m,n)$\in$B}w_\text{m} w_\text{n}.
\end{equation}
Here, $W_\text{A} = \sum_\text{(m,n)$\in$A} w_\text{m} w_\text{n}$ and $w_\text{m}, w_\text{n}$ are the weights for each absorber and are redshift dependent such that $w \propto (1+z)^\gamma$. $(m,n)\in A$ refers to separations calculated using the reconstructed redshift of the absorber pair and $(m,n)\in B$ refers to separations computed using the redshifts $z_\text{m}$ and $z_\text{n}$ of the absorber pair $(m,n)$. 

In the fits to the data each metal species has its own individual bias parameters $(b, \beta)$. Since the correlations are only visible in (\rp, \rt) bins that correspond to small physical separations and the amplitudes for the \texttt{SiII} and \texttt{SiIII} metal species are determined by the excess correlations, we do not have enough signal to determine the $(b, \beta)$ parameters separately. As in previous works (\cite{dmdb2020, dmdb2017}), we fix $\beta = 0.5$ for all metal species. Since we are using the auto-correlation from \cite{calum2023} we include correlations from the four metal species given in table \ref{table:metals} in the correlations.

%transverse proximity effect
The last term in eq.~\ref{eq:model_components} is due to the transverse proximity effect. In the area surrounding a quasar, there is significantly less \lya absorption because the radiation emitted from the quasar dominates over the UV background and increases the ionization fraction of the surrounding gas, thus making it more transparent to \lya photons. 
The \lya absorption of a background quasar will decrease in strength at the redshift of the foreground quasar when the \lya forest is close to the quasar line of sight. This will therefore affect the correlation between quasars and the \lya forest. As in \cite{calum2023}, we model this effect in the correlation with 
\begin{equation}
    \label{eq:TPE}
    \xi^\text{TP} = \frac{\xi^\text{TP}_0}{r^2}\exp{\left[\frac{-r}{\lambda_{\text{UV}}}\right]},
\end{equation}
where $r$ is the comoving separation in units of \mpc, $\lambda_{\text{UV}} = $\mpch{300} \cite{rudie2013}, and $\xi_0^{TP}$ is an amplitude that will be fit.

Systematic errors on the measurement of the quasar redshift can cause a shift in the model estimation of the correlation function in \rp. To account for this, we allow for a shift of the absorber-quasar separation in the \rp direction 
\begin{equation}
    \label{eq:drp}
    \Delta r_\parallel = r_{\parallel,\text{true}} - r_{\parallel,\text{measured}}.
\end{equation}
The shift is mostly due to systematic quasar redshift errors. However, any asymmetries in \rp will also affect \drp. The model of the cross-correlation function is not symmetric about \rp due to the contribution of metal absorptions and the variation of mean redshift with \rp. The continuum-fitting distortion also introduces an asymmetry in \rp. \cite{lepori2020} has showed that while \drp is sensitive to relativistic effects, they are small. We do not model any relativistic effects in this work as \cite{blomqvist2019} showed that they are partially degenerate with \drp.

\section{Validation with Mock Data}
\label{sec:mocks}

To ensure that the choice of bin width does not introduce any biases to our analysis, and especially to the two redshift error parameters \drp and $\sigma_\text{v}$, we run our tests using mock data for three different cross-correlation bin widths: \mpch{1}, \mpch{2}, and \mpch{4}. We follow the same method described in section \ref{sec:method}, however we do not need to perform the calibration step on mock data (section \ref{subsec:contfit}) during the continuum fitting process. For the cross-correlation, we still restrict \rp, \rt $\in$ \mpch{[-80,80]}. For the \mpch{2} bin width, this gives 80 bins along the line of sight, 40 bins across the line of sight and N = 3,200 bins in total. For the \mpch{4} bin width, this gives 40 bins along the line of sight, 20 bins across the line of sight, and N = 800 bins in total. Though the auto-correlation does not help to constrain redshift errors we include it to help break degeneracies with other parameters. We follow the setup and method from \cite{calum2023} for the auto-correlation. We restrict \rp, \rt $\in [0,200]$\mpc using 50 bins with a 4 \mpc bin width, which gives 2500 total bins. The free and fixed parameters are described in table \ref{table:modelparams} in appendix \ref{sec:appfitpars}. 

In this work we use the mocks described in \cite{hiram2023}. There are 10 realizations of Saclay \cite{saclay} mocks that contain signal from the \lya forest as well as contamination from DLAs, BALs, and metals, which most closely resembles the data from EDR+M2. Like in our baseline analysis, we mask the BAL and DLA regions during the continuum fitting process, and we remove any BAL quasars from the tracer quasar catalog. Each mock has approximately 118,000 objects in the catalog after BAL removal.  

\begin{figure}
    \centering
    \includegraphics[width=15cm]{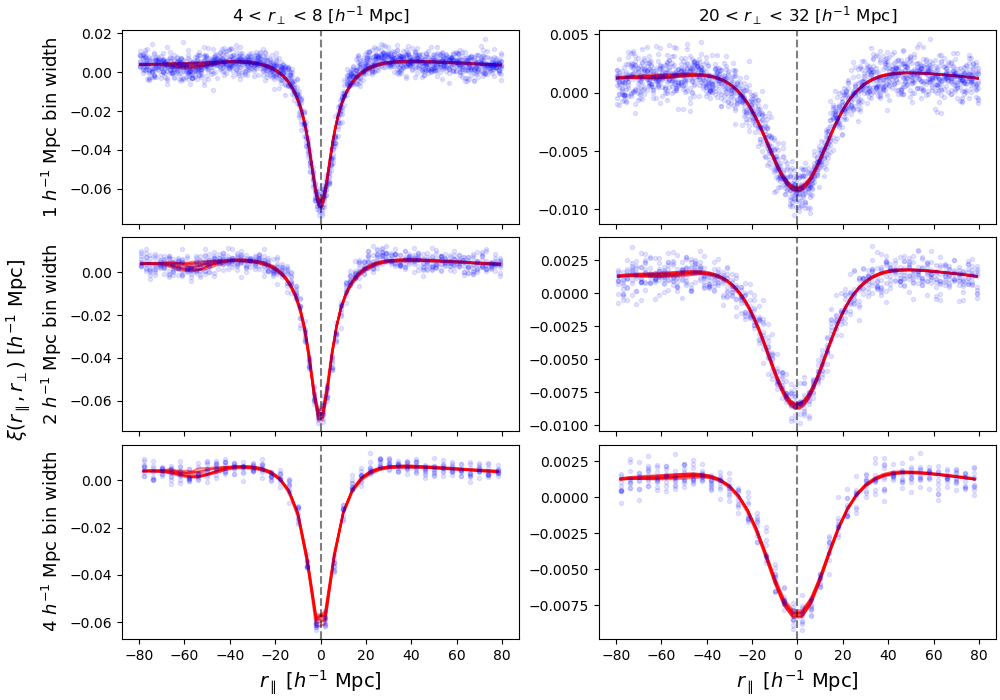}
    \caption{Results for 10 different realizations of mocks comparing the correlation function to the best-fit model. The data (blue points) and model (red lines) are averaged across \rt $\in$ [4,8] (left column) and \rt $\in$ [20,32] (right column). We compare the different bin widths studied in this analysis with \mpch{1} on the top row, \mpch{2} in the middle, and \mpch{4} on the bottom row. The dashed grey line represents \rp = 0.}
    \label{fig:mocksmodel}
\end{figure}

We present the results of the cross-correlation and best-fit model for each mock for two \rt slices (\rt $\in$ [4,8]\mpc and \rt $\in$ [20,32]\mpc) in figure~\ref{fig:mocksmodel}. The data and model are averaged across the \rt slices for each mock. There are 4 times fewer points in the bottom row of figure \ref{fig:mocksmodel} compared to the top row due to the bin width being 4 times larger. There is a bump in the correlations around \rp $\approx$ \mpch{-60} that likely corresponds to the effects from SiII $\lambda$1193 and $\lambda$1190 metal lines, which appear in the correlations as given in table \ref{table:metals}. This bump is not as prominent in the larger \rt slices as the metals have the largest effect at \rt = 0. There does not appear to be a shift from 0 in \rp, as the center of the peaks corresponds with the dashed grey line at \rp = 0. 

\begin{figure}
    \centering
    \includegraphics[width=14cm]{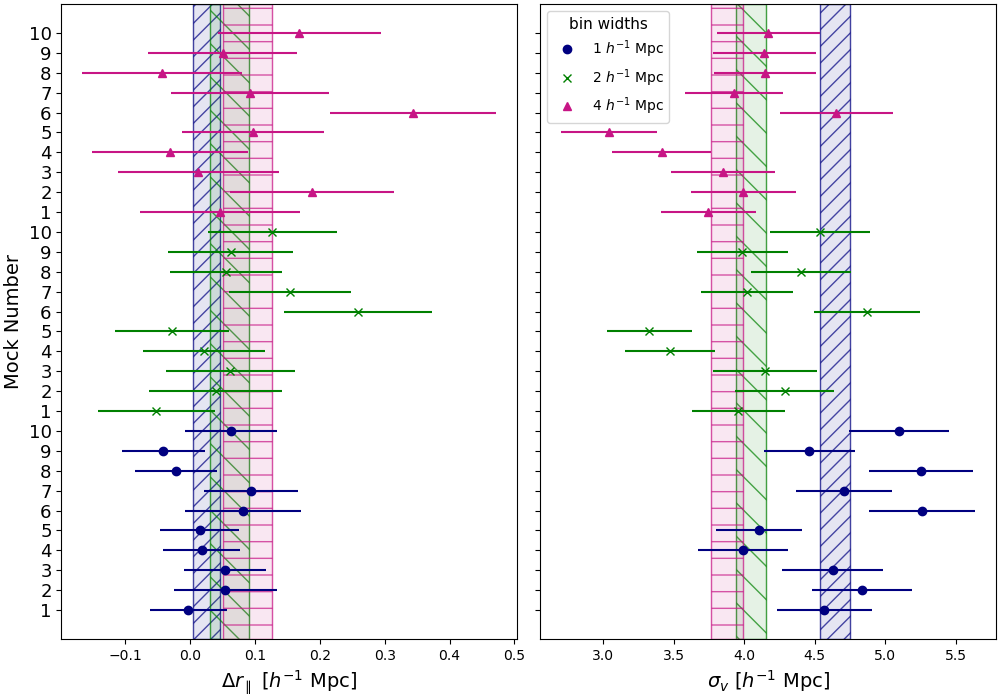}
    \caption{Results for 10 different realizations of mocks comparing the redshift error parameters, \drp (left) and $\sigma_\text{v}$ (right) for the three different bin widths being studied: \mpch{1} (blue dots), \mpch{2} (green x's), and \mpch{4} (pink triangles). The hatched shaded regions are showing the weighted average value + error for all 10 mocks.}
    \label{fig:zerrparsmocks}
\end{figure}

The two parameters of interest in our model that describe redshift errors are \drp and $\sigma_\text{v}$. \drp describes the shift of the cross-correlation due to systematic redshift errors in quasars along the \rp direction. $\sigma_\text{v}$ measures the radial velocity smearing due to random redshift errors and non-linear velocities. This is often represented as the rms velocity dispersion from a Lorentz damping factor \cite{pw2009, dmdb2017}. Here, a smaller value for $\sigma_\text{v}$ means the redshift measurements are more accurate.
 
The mocks studied here do not have any systematic errors added to the quasar redshift so we expect \drp $\sim 0$. The true value of $\sigma_\text{v}$ is known and we do not expect it to be 0, as the mocks include non-linear effects. The results for these parameters for each mock and each bin width tested are shown in figure \ref{fig:zerrparsmocks}. The weighted average across all mocks is shown by the hatched shaded regions. For the \mpch{1} bin width: \drp = $0.039 \pm 0.021$\mpc and $\sigma_\text{v} = 4.64 \pm 0.11$\mpc. For the \mpch{2} bin width: \drp = $0.079 \pm 0.030$\mpc and $\sigma_\text{v} = 4.05 \pm 0.11$\mpc. For the \mpch{4} bin width: \drp = $0.10 \pm 0.04$\mpc and $\sigma_\text{v} = 3.88 \pm 0.11$\mpc. The $\chi^2$ values averaged over all mocks are $11908/(11644 - 12) = 1.023$ for the \mpch{1} bin width, $4182/(4102 - 12) = 1.022$ for the \mpch{2} bin width, and $2230/(2222 - 12) = 1.009$ for the \mpch{4} bin width, which all indicate a good fit.

Results for \drp with different bin widths are in good agreement with each other. They are also consistent with the true value of \drp = 0, although we cannot rule out a small systematic uncertainty as large as \mpch{0.1}. The results for $\sigma_\text{v}$ are not consistent across the different bin widths. This is likely because the true value of $\sigma_\text{v}$ in the mocks is too small to measure with the \mpch{4} bin width. The mocks have a resolution of \mpch{2.5} - which means we would need to additionally model the effect of resolution for the 1 and \mpch{2} bin widths. To test that we can recover the true value for $\sigma_\text{v}$, regardless of bin width, we add Gaussian random errors with $\sigma$ = \mpch{10} to the redshifts in the quasar catalogs. We then perform the continuum fitting process and calculate and fit the correlations for each mock. We take the weighted average across all 10 mocks and find $\sigma_\text{v} = 12.93 \pm 0.25$\mpc, $12.12 \pm 0.25$\mpc, and $11.14 \pm 0.24$\mpc for the 1, 2, and \mpch{4} bins, respectively. The increased values show that we can recover the correct value for $\sigma_\text{v}$.

Overall, the \mpch{1} bin widths do not appear to cause any biases in \drp larger than \mpch{0.1} when compared with the \mpch{2} and \mpch{4} bin widths. Since the larger \mpch{4} bins will smooth out any small-scale features from systematic redshift errors, we will present results on the EDR+M2 data in section \ref{sec:results} using a \mpch{1} bin width.

\section{Results}
\label{sec:results}
In this section we present the results of this analysis using the auto-correlation from \cite{calum2023} and the cross-correlation measured from the data described in section~\ref{sec:data} and using the methodology described in section~\ref{sec:method}. When necessary, we re-run the auto-correlation following the method in \cite{calum2023}. We present the cross-correlation and best-fit model for the baseline configuration in section~\ref{subsec:corrfit}. The baseline configuration masks both BAL and DLA regions of the quasar spectra during the continuum fitting process, and BALs are removed from the tracer quasar catalog. We explore the dependence on quasar redshift in section~\ref{subsec:zdep} and we discuss the effect of including and using updated redshifts for BALs in section~\ref{subsec:bal}. All correlations and fits are calculated and measured using a \mpch{1} bin width. Again, in this study we adopt the flat $\Lambda$CDM cosmology of Planck2018 \cite{planck}.

\subsection{Cross-correlation and Baseline Fit Results}
\label{subsec:corrfit}

\begin{figure}
    \centering
    \includegraphics[width = \textwidth]{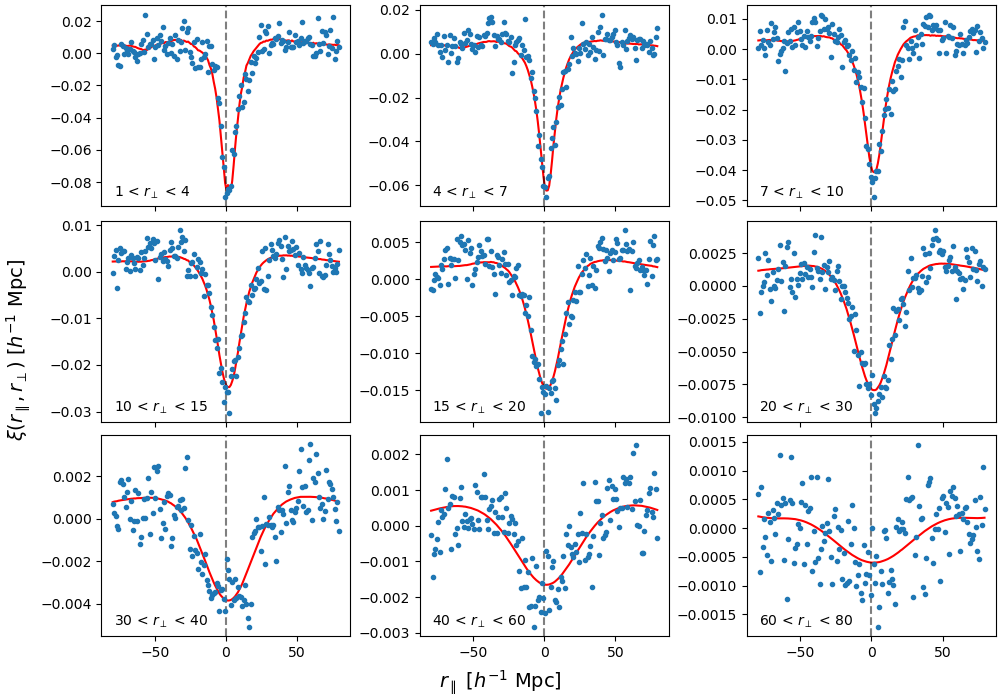}
    \caption{Measured \lya forest and quasar cross-correlation (blue points) and resulting best-fit model for the baseline fit (red line). Both the correlation and the model are averaged in slices of \rt and the \rt slices (in \mpc) are shown in the lower left of each subplot. The dashed grey line represents \rp = 0. Seen in all \rt slices is a shift of the correlation and model in the positive \rp direction, which is consistent with our results for \drp.}
    \label{fig:xcorr}
\end{figure}

The measured \lya forest and quasar cross-correlation and the result for the baseline fit are presented in figure~\ref{fig:xcorr}. We show the average of the correlation and model in each \rt slice listed. The correlations have a lower amplitude for larger \rt slices and therefore appear noisier. The model for the baseline fit is calculated jointly using the auto- and cross-correlations and has 13 free parameters. The values for the fixed parameters used in this model are presented in table \ref{table:modelparams} in Appendix \ref{sec:appfitpars}. 

\begin{table}[]
    \centering
    \begin{tabular}{c|ccc}
        Parameter & Baseline & BAL & ZBAL \\ \hline \hline
        No. quasars & 106,861 & 128,026 & 127,989 \\ \hline
        
        $b_{\text{Ly}\alpha}$ & $-0.106 \pm -0.023$ & $-0.053 \pm 0.011$ & $-0.048 \pm 0.009$ \\
        
        $\beta_{\text{Ly}\alpha}$ & $1.8 \pm 0.4$ & $3.7 \pm 0.7$ & $4.0 \pm 0.7$ \\
        
        $10^3b_{\eta,\text{SiII(1190)}}$ & $-2.8 \pm 0.9$ & $-2.6 \pm 0.8$ & $-2.1 \pm 1.0$ \\
        
        $10^3b_{\eta,\text{SiII(1193)}}$ & $-0.3 \pm 1.0$ & $-0.0 \pm 1.3$ & $-0.0 \pm 1.0$ \\
        
        $10^3b_{\eta,\text{SiIII(1207)}}$ & $-2.6 \pm 0.9$ & $-1.9 \pm 0.9$ & $-1.2 \pm 0.9$ \\

        $10^3b_{\eta,\text{SiII(1260)}}$ & $-2.7 \pm 1.2$ & $-3.9 \pm 1.6$ & $-4.9 \pm 1.7$ \\
        
        $b_\text{HCD}$ & $-0.051 \pm 0.008$ & $-0.054 \pm 0.010$ & $-0.052 \pm 0.011$ \\
        
        $b_\text{q}$ & $3.1 \pm 0.4$ & $4.1 \pm 0.4$ & $4.4 \pm 0.5$ \\
        
        \drp ($h^{-1}$Mpc) & $-1.94 \pm 0.15$ & $-1.84 \pm 0.13$ & $-1.90 \pm 0.11$ \\
        
        $\sigma_\text{v}$ ($h^{-1}$Mpc) & $5.8 \pm 0.5$ & $7.3 \pm 0.6$ & $7.0 \pm 0.6$ \\
        
        $\xi_0^\text{TP}$ & $0.29 \pm 0.05$ & $0.08 \pm 0.024$ & $0.082 \pm 0.023$  \\ 

        $10^4A_\text{inst}$ & $2.28 \pm 0.31$ & $1.95 \pm 0.27$ & $1.90 \pm 0.27$ \\
        
        \hline 

        $\chi^2$ & 12051.4 & 12121.1 & 12058.6 \\
        
        $\chi^2$/dof & 1.036 & 1.042 & 1.037 \\
        
        $z_\text{eff}$ & 2.34 & 2.34 & 2.35 \\
   
    \end{tabular}
    \caption{Results of the fits for the baseline, BAL, and the ZBAL configurations. The number of quasars above $z = 1.88$ in each catalog is given. In all fits the BAO parameters are fixed at 1.0. The rest of the parameters are allowed to float in each fit and are described in Table \ref{table:modelparams} in appendix \ref{sec:appfitpars}. Each fit has 11,644 data bins, 13 free parameters and we also list the effective redshift, $z_\text{eff}$. Since each fit is a joint fit of the auto- and cross-correlations, we list the $\chi^2$ and reduced $\chi^2$ ($\chi^2$/dof) values given for the joint fit.}
    \label{table:params}
\end{table}

We present the results of the baseline fit in the first column in table \ref{table:params}. The baseline fit has a joint $\chi^2$ value that corresponds to essentially zero probability. This is likely due to our model failing at small scale separations. We re-fit our model, increasing the minimum separation to \mpch{5} and \mpch{10} to check the results of the $\chi^2$ values. We found a very minimal improvement in $\chi^2$ values and a worsening of \drp values. We therefore decide to keep the minimum separation in our fit at \mpch{0} and leave for future studies any further tests of the model on $\chi^2$ values.

We find \drp $= -1.94 \pm 0.15$\mpc and $\sigma_\text{v} = 5.8 \pm 0.5$\mpc. Comparing these results to table 1 of \cite{calum2023}, we find that $\sigma_\text{v}$ is within 1.0$\sigma$ from the value reported in \cite{calum2023}. Our result for \drp is also consistent within 1.5$\sigma$ with \cite{calum2023}, while our fit value for $\xi_0^\text{TP}$ is 2.0$\sigma$ with respect to the results reported in \cite{calum2023}. These differences are likely due to our smaller bin width, and the different ranges in \rp and \rt used.

A negative value for \drp means the measured cross-correlation is shifted in the positive \rp direction, which is seen in all 9 \rt slices in figure \ref{fig:xcorr}. Based on our definition of a positive separation, i.e, when the quasar is in front of the \lya pixel, this indicates that the estimated quasar redshifts are less than the true quasar redshifts.

Our initial thought for this discrepancy is the disappearance of the \texttt{MgII} emission line from quasar spectra above redshift $z \approx 2.5$. The \texttt{MgII} line, a reliable indicator of systemic redshift \cite{hewettwild10, afr2013}, is located beyond the observable range of the DESI spectrographs at these redshifts. We therefore rely on the high ionization broad lines, which often display significant velocity shifting \cite{shen2016}, to obtain redshift estimates at $z > 2.5$. This is also suggested by \cite{edmond}, who report a ``kink" at $z \sim 2.5$ when comparing the redshift estimates from DESI to those from SDSS (their figure 16). However, this is unlikely the cause of the ``kink" due to the observable range of the BOSS spectrographs being very similar to DESI \cite{boss}, as eBOSS would lose access to the \texttt{MgII} line at similar redshifts to DESI.

The apparent underestimation of DESI redshifts persisted when updating the quasar templates in Redrock for processing the DESI year one quasar sample, as they did not account for the \lya optical depth \cite{brodzeller2023}. In section \ref{sec:zdep}, we discuss the improvements in redshift performance for DESI year 1 data at $z > 2$ when using an updated version of Redrock and quasar templates that properly model \lya optical~depth.

\subsection{Dependence on Quasar Redshift}
\label{subsec:zdep}

\begin{table}[]
    \centering
    \begin{tabular}{c|cccc}
        Parameter & z1 & z2 & z3 & z4 \\ \hline \hline
        No. quasars & 41,688 & 27,958 & 17,407 & 19,808  \\
        Redshift range & $z \le 2.2$ & $2.2 < z \le 2.5$ & $2.5 < z \le 2.8$ & $z > 2.8$ \\ \hline 
        
        $b_{\text{Ly}\alpha}$ & $-0.085 \pm 0.010$ & $-0.117 \pm 0.009$ & $-0.160 \pm 0.011$ & $-0.216 \pm 0.015$ \\
        
        $\beta_{\text{Ly}\alpha}$ & $1.77 \pm 0.21$ & $1.63 \pm 0.17$ & $1.48 \pm 0.16$ & $1.45 \pm 0.15$ \\
        
        $10^3b_{\eta,\text{SiII(1190)}}$ & $-3.6 \pm 1.0$ & $-2.6 \pm 1.0$ & $-3.2 \pm 1.1$ & $-3.4 \pm 1.2$ \\
        
        $10^3b_{\eta,\text{SiII(1193)}}$ & $-0.9 \pm 0.9$ & $-1.3 \pm 1.0$ & $-1.4 \pm 1.1$ & $-1.5 \pm 1.2$ \\
        
        $10^3b_{\eta,\text{SiIII(1207)}}$ & $-3.9 \pm 0.9$ & $-2.6 \pm 0.9$ & $-3.8 \pm 1.0$ & $-4.2 \pm 1.2$ \\

        $10^3b_{\eta,\text{SiII(1260)}}$ & $-2.8 \pm 1.1$ & $-2.6 \pm 1.1$ & $-2.9 \pm 1.2$ & $-3.1 \pm 1.3$ \\
        
        $b_\text{HCD}$ & $-0.037 \pm 0.007$ & $-0.049 \pm 0.008$ & $-0.053 \pm 0.011$ & $-0.067 \pm 0.015$ \\
        
        $b_\text{q}$ & $2.73 \pm 0.29$ & $2.88 \pm 0.14$ & $3.20 \pm 0.23$ & $4.1 \pm 0.5$ \\
        
        \drp ($h^{-1}$Mpc) & $-1.07 \pm 0.17$ & $-1.99 \pm 0.13$ & $-3.5 \pm 0.4$ & $-3.4 \pm 1.0$ \\
        
        $\sigma_\text{v}$ ($h^{-1}$Mpc) & $5.4 \pm 0.7$ & $5.5 \pm 0.6$ & $6.3 \pm 1.0$ & $8.1 \pm 1.8$ \\
        
        $\xi_0^\text{TP}$ & $0.13 \pm 0.04$ & $0.18 \pm 0.04$ & $0.51 \pm 0.13$ & $0.8 \pm 0.4$ \\ 
        
        $10^4A_\text{inst}$ & $2.29 \pm 0.29$ & $2.38 \pm 0.27$ & $2.42 \pm 0.27$ & $2.42 \pm 0.27$ \\
        
        \hline 

        $\chi^2$ & 12010.9 & 11978.3 & 11980.5 & 12248.9 \\
        
        $\chi^2$/dof & 1.033 & 1.030 & 1.030 & 1.053 \\
        
        $z_\text{eff}$ & 2.09 & 2.33 & 2.62 & 2.99  \\
         
    \end{tabular}
    \caption{Results of the fits for the four redshift evolution bins. The number of quasars in each bin is given, along with the redshift range of each bin. In all fits the BAO parameters are fixed at 1.0. The rest of the parameters are allowed to float in each fit and are described in table \ref{table:modelparams} in appendix \ref{sec:appfitpars}. Each fit has 11,644 data bins, 13 free parameters and we also list the effective redshift, $z_\text{eff}$. Since each fit is a joint fit of the auto- and cross-correlations, we list the $\chi^2$ and reduced $\chi^2$ ($\chi^2$/dof) values given for the joint fit.}
    \label{table:zrange}
\end{table}

We test and study here the evolution and dependence of the parameters \drp and $\sigma_\text{v}$ on quasar redshift to investigate more in depth the cause of the measured biases. We split the tracer quasar sample described in section \ref{sec:data} into four redshift bins: z1: $z \le 2.2$, z2: $2.2 < z \le 2.5$, z3: $2.5 < z \le 2.8$, and z4: $z > 2.8$. As a reminder, quasars with redshift $z < 1.88$ will not contribute to the correlation as the separation will be larger than our maximum separation of \mpch{80}. We use the same sample of \lya forests as the baseline, but we create new quasar catalogs with approximately 41,000 objects in the first catalog, 28,000 objects in the second, 17,000 objects in the third, and 19,000 objects in the fourth. We then re-calculate the cross-correlation and fit for each redshift bin following the methods in section~\ref{sec:method}. Since the auto-correlation only depends on the \lya forest sample, we do not need to re-compute it and can use the auto-correlation from \cite{calum2023} for each redshift bin when computing the joint fit.

\begin{figure}
    \centering
    \includegraphics[width=\textwidth]{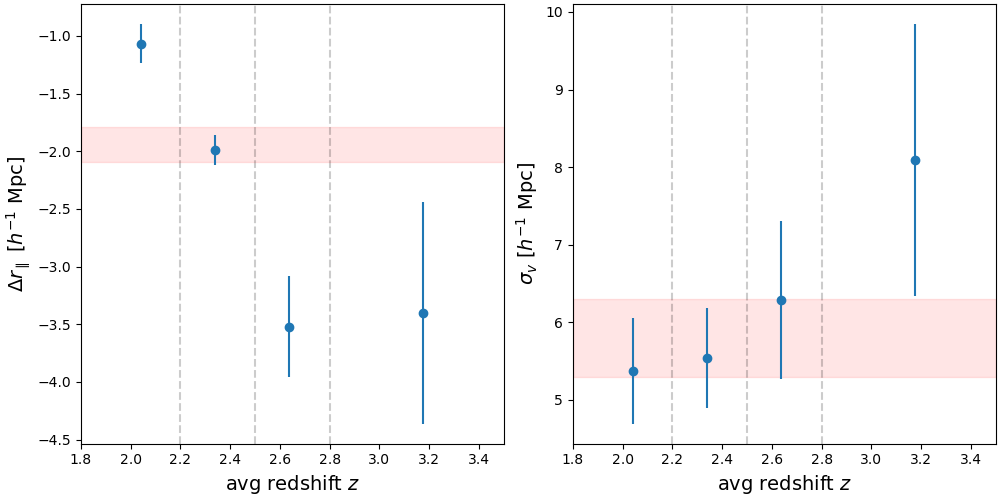}
    \caption{Evolution of \drp and $\sigma_\text{v}$ with redshift for four different redshift bins (blue points). The bin edges at z = 2.2, 2.5, and 2.8 are marked by the dashed grey lines. The red shaded region denotes the baseline value for each parameter: \drp = $-1.94 \pm 0.15$\mpc and $\sigma_\text{v} = 5.8 \pm 0.5$\mpc.}
    \label{fig:zevol}
\end{figure}

The fit results for each redshift bin are given in table~\ref{table:zrange}. Each fit has the same configuration as the baseline fit presented in the first column of table \ref{table:params}. We plot these results vs the average redshift in each bin in figure~\ref{fig:zevol}, highlighting the baseline value in the shaded region. We find a strong evolution with redshift for \drp with the value becoming more negative as the average redshift of the bin increases for the first three bins. However, the fourth redshift bin does not clearly follow this trend. Though this bin does not have the least amount of quasars, the quasars in this bin are at a higher redshift and are generally fainter and have lower signal to noise. The fit for this redshift bin is also the worst of the four with a $\chi^2 = 1.053$, which corresponds to near-zero probability given the 11,631 degrees of freedom in the fit. Despite this, the overall trend indicates that the quasar redshifts are being underestimated by a larger amount as the quasar redshift increases, which is consistent with the quasar templates not accounting for the \lya forest optical depth. If the loss of the \texttt{MgII} emission line in the quasar spectra was the dominant contribution to the redshift errors, we would expect a sudden jump in \drp and $\sigma_\text{v}$ as a function of $z$ around $z = 2.5$. Such a feature is not observed, so the loss of \texttt{MgII} is unlikely to be the culprit.

The measured evolution of $\sigma_\texttt{v}$ as a function of $z$ is much less striking than that of \drp. Still, there is an indication for a mild increase of $\sigma_\texttt{v}$ for higher redshift quasars, consistent with less precision in the measured redshift.

\subsection{Impact of BALs and Updating BAL Quasar Redshifts}
\label{subsec:bal}

\begin{figure}
    \centering
    \includegraphics[width = \textwidth]{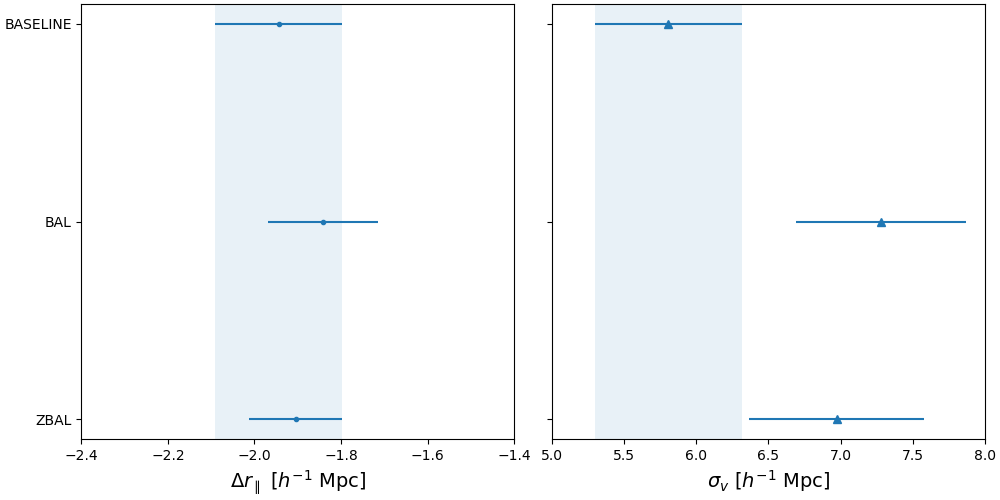}
    \caption{Results on \drp (left, blue points) and $\sigma_\text{v}$ (right, blue triangles) for the three different configurations presented in this work: baseline, BAL, and ZBAL. The shaded area represents the value of the parameters from the baseline fit.}
    \label{fig:lep}
\end{figure}

BAL systems can impact the shape of quasar emission lines and introduce shifts in the estimated quasar redshift, up to several hundred km s$^{-1}$ \cite{bal1}. BAL quasars often have larger redshift errors compared to quasars with no BAL features because the BAL features impact the blue side of emission lines \cite{garcia2023}. This is true in DESI, where Redrock tends to overestimate redshifts for BAL quasars \cite{garcia2023}. The current pipeline used for most \lya studies, \texttt{Picca}, has the option to mask out BAL regions during the continuum fitting process, but does not update the quasar redshift when these masks are applied so any systematic errors on the quasar redshift will still be present. \cite{bal1} shows that when BAL regions of a quasar spectrum are masked and new redshifts are obtained, those new redshifts shift by 240 km s$^{-1}$ on average. We explore in this section the effect of including the BAL quasars in the tracer catalog as well as using the updated BAL redshifts from \cite{bal1} once the BAL regions of the quasar spectra are masked. 

We first test if including BAL quasars in the tracer quasar catalog will have an effect on \drp or $\sigma_\text{v}$. For this test we use the same \lya forest sample as \cite{cesar} which is the same as our baseline. The tracer quasar catalog is the same as described in section \ref{subsec:qso} except the BALs are not removed. It consists of 128,026 quasars with redshift $z > 1.88$. We use the auto-correlation from \cite{calum2023} and follow the same method described in section \ref{sec:method} to measure and fit the cross-correlation. We find \drp = $-1.84 \pm 0.13$~\mpc and $\sigma_\text{v} = 7.3 \pm 0.6$~\mpc. These results, along with those from the baseline configuration, are shown in figure \ref{fig:lep} and presented in table \ref{table:params} in the BAL column. The shift of \drp in the positive direction would seem to indicate that including BAL quasars in the tracer catalog reduces the quasar redshift errors. However, this is actually due to BAL quasars typically having overestimated redshifts, as the BAL features tend to impact the blue side of the emission lines \cite{garcia2023}. Any overestimation of quasar redshifts would shift \drp in the positive direction. This is the opposite of what we observe when the BAL quasars are excluded, where the shift in \drp is negative due to the underestimation of redshifts. It is likely that the inclusion of the BAL quasars partially compensates for the negative shift.

To test if updating the BAL quasar redshifts after the BAL regions are masked improves the \drp parameter, we use the quasar catalog from \cite{bal1}. This is the same catalog as described above, but with updated redshifts for the BAL quasars, of which 28,185 BAL quasars have an updated redshift. Because the redshifts are updated, the \lya forest sample changes slightly, both in the number of forests and the redshift distribution of pixels. While the effect is small, we make the choice to not use the \lya forest sample described in \cite{cesar}. We follow the method in \cite{cesar} and in section \ref{sec:method} for the continuum fitting process and our final \lya sample consists of 88,432 forests. We use the tracer catalog with updated redshifts and follow the method described in section \ref{sec:method} to measure and fit the correlations. We can no longer use the auto-correlation from \cite{calum2023}, so we measure those following the method in \cite{calum2023}. The results are shown in figure \ref{fig:lep} and in table \ref{table:params} in the ZBAL column. We find \drp~=~$-1.90 \pm 0.11$~\mpc and $\sigma_\text{v} = 7.0 \pm 0.6$~\mpc. The value of \drp has shifted in the negative direction and is now equal ($0.35\sigma$) to the baseline value. This negative shift is consistent with what is found in \cite{bal1}, which found that the redshifts of BAL quasars after masking were shifted to the blue. 

In both configurations when BAL quasars are added in at the catalog level the values of $b_{\text{Ly}\alpha}$ and $\beta_{\text{Ly}\alpha}$ change by a factor of two compared to the baseline. These parameters are highly correlated with each other and with $b_\text{q}$ ($\rho \sim 0.9$). By adding the BAL quasars at the catalog level we are changing the value of $b_\text{q}$, so it is not unexpected that $b_{\text{Ly}\alpha}$ and $\beta_{\text{Ly}\alpha}$ also change. 

This shows that BAL quasars can affect the correlations when they are included in the tracer catalog and their redshifts are not updated after the BAL regions are masked. When the redshifts are updated the inclusion of BAL quasars in the tracer catalog has no effect on~\drp.

\section{Addressing the Redshift Dependency}
\label{sec:zdep}
The redshift dependency in \drp is also reported in the year one quasar sample by \cite{brodzeller2023}, who suggest the bias could be mitigated through proper modeling of the mean transmission of \lya in the spectral templates used for redshift estimation. We define the mean transmitted flux fraction, $\overline{F}$, as 
\begin{equation}
\label{eq:transmittedFlux}
    \overline{F}(z) = e^{-\tau_{\text{eff},\alpha} (z)},
\end{equation}
where
\begin{equation}
\label{eq:effOpticalDepth}
    \tau_{\text{eff},\alpha} = \tau_{0} (1 + z)^{\gamma}
\end{equation}
is the effective optical depth of the \lya transition, $\tau_0 = (5.54 \pm 0.64)\times 10^{-3}$ and $\gamma = 3.182 \pm 0.074$ \cite{kamble2020}. $\overline{F}$ depends on $\tau_{\text{eff},\alpha}$ at $z=\frac{\lambda_\text{obs}-\lambda_{\text{Ly}\alpha}}{\lambda_{\text{Ly}\alpha}}$ according to eq.~\ref{eq:transmittedFlux}. $\tau_{\text{eff},\alpha}$ evolves with redshift, as parameterized by eq.~\ref{eq:effOpticalDepth} \citep[e.g.,][]{kim2007,calura2012,kamble2020}, implying greater overall flux suppression in the \lya forest region of quasar spectra at higher redshifts. The BOSS quasar templates used for EDR and the DESI year 1 quasar templates were derived from samples of quasar spectra that were not corrected for the optical depth of \lya photons. On the other hand, a real-time correction for this flux suppression was being applied in Redrock during redshift fitting, leading to an over-suppression of template flux relative to what was expected at $\lambda_\text{RF} < 1215.67$\AA. This resulted in a bias that increases with the amount of suppression (i.e. redshift) at $z > 2$. 

We update the DESI year one quasar templates with the \lya effective optical depth model from \cite{kamble2020} and test whether the observed redshift bias is mitigated. The year one quasar templates consist of two eigenspectra sets which are trained to classify separate, yet overlapping redshift ranges: $0.05<z<1.6$ and $1.4<z<7.0$, dubbed \texttt{LOZ} and \texttt{HIZ}, respectively. As the \lya forest is only observed in DESI quasar spectra with $z>1.95$, only the \texttt{HIZ} templates require modification. A new version of the \texttt{HIZ} templates are derived from the same sample as in \cite{brodzeller2023}, but the individual spectra are corrected for suppression from \lya effective optical depth at $\lambda_{RF} < 1215.67$\AA. The \lya optical depth model in Redrock was then updated accordingly.\footnote{\url{https://github.com/desihub/redrock/releases/tag/0.17.9}} Our updated \texttt{HIZ} templates can be found on Github.\footnote{\url{https://github.com/desihub/redrock-templates/releases/tag/0.8.1}}

\begin{figure}
    \centering
    \includegraphics[width = \textwidth]{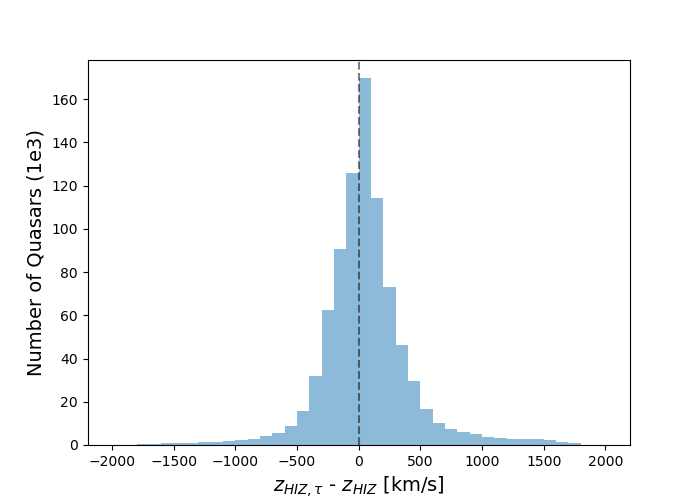}
    \caption{Distribution of the difference in redshift (in km~s$^{-1}$) for the quasars in the year 1 catalog between the redshifts measured with the updated \texttt{HIZ} templates ($z_{\texttt{HIZ},\tau}$) versus the redshifts measured from the templates without the \lya optical depth correction ($z_\texttt{HIZ}$).}
    \label{fig:zdep}
\end{figure}

We use the updated \texttt{HIZ} templates with modified Redrock to refine the redshifts of the DESI year 1 quasar sample at $z > 1.6$. This redshift is defined by the upper limit of the redshift coverage for the \texttt{LOZ} templates, though we note substantial impact is only expected at $z > 1.95$. We impose a prior of $\Delta v = \pm 3000$ km~s$^{-1}$ from the original redshift estimate, as the \texttt{HIZ} redshifts have been shown to be approximately correct within a few tens to hundreds of km~s$^{-1}$, on average. Approximately 850,000 objects had a change in redshift when using the updated \texttt{HIZ} templates, and we show the distribution of those changes in figure \ref{fig:zdep}.

\begin{figure}
    \centering
    \includegraphics[width=\textwidth]{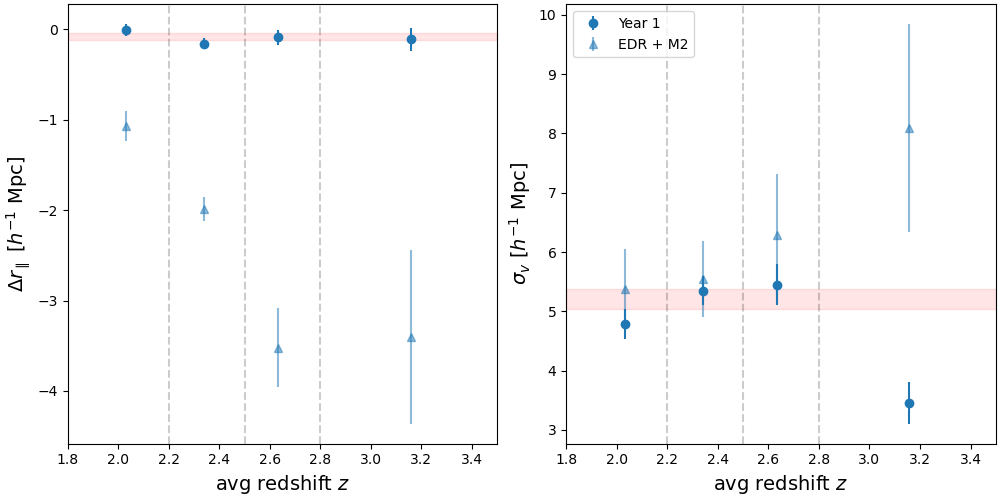}
    \caption{Evolution of \drp and $\sigma_\text{v}$ with redshift using the updated quasar templates for the four different redshift bins (blue points). The results with the old templates are shown in light blue triangles for comparison. The bin edges at z = 2.2, 2.5, and 2.8 are marked by the dashed grey lines. The red shaded region denotes the baseline value for each parameter: \drp~=~$-0.08 \pm 0.04$\mpc and $\sigma_\text{v} = 5.21 \pm 0.17$\mpc.
     }
    \label{fig:zevol_iron}
\end{figure}

To test that the redshift bias shown in \drp is mitigated we perform our analysis again using the updated quasar templates. We use the year 1 \lya forest sample and tracer quasar catalog from \cite{desikp6}. To be consistent with our baseline analysis we remove any BAL quasars from the tracer catalog. There are 510,000 quasars in the tracer catalog with redshift $z > 1.88$. We also split the tracer quasar sample to test for any evolution with redshift with the same cuts as in section \ref{subsec:zdep}: $z = 2.2$, 2.5, and 2.8. We measure and fit the correlations using the same method described in section \ref{sec:method}, and we show the results in table \ref{table:iron} and figure~\ref{fig:zevol_iron}.

\begin{table}[]
    \begin{adjustbox}{width=\columnwidth}
    \centering
    \begin{tabular}{c|ccccc}
        Parameter & baseline & z1 & z2 & z3 & z4 \\ \hline \hline 
        No. quasars & 510,501 & 201,290 & 138,364 & 82,710 & 88,137 \\
        Redshift range & - & $z \le 2.2$ & $2.2 < z \le 2.5$ & $2.5 < z \le 2.8$ & $z > 2.8$ \\ \hline 
        
        $b_{\text{Ly}\alpha}$ & $-0.0543 \pm 0.0035$ & $-0.049 \pm 0.005$ & $-0.055 \pm 0.005$ & $-0.060 \pm 0.006$ & $-0.162 \pm 0.020$ \\
        
        $\beta_{\text{Ly}\alpha}$ & $3.78 \pm 0.26$ & $3.3 \pm 0.4$ & $3.8 \pm 0.4$ & $4.5 \pm 0.5$ & $2.01 \pm 0.28$ \\
        
        $10^3b_{\eta,\text{SiII(1190)}}$ & $-2.5 \pm 0.4$ & $-2.6 \pm 0.4$ & $-2.8 \pm 0.4$ & $-3.9 \pm 0.5$ & $-3.4 \pm 0.5$ \\
        
        $10^3b_{\eta,\text{SiII(1193)}}$ & $-1.34 \pm 0.35$ & $-1.3 \pm 0.4$ & $-2.2 \pm 0.4$ & $-2.0 \pm 0.5$ & $-2.2 \pm 0.5$ \\
        
        $10^3b_{\eta,\text{SiIII(1207)}}$ & $-4.50 \pm 0.31$ & $-5.0 \pm 0.4$ & $-5.0 \pm 0.4$ & $-6.0 \pm 0.5$ & $-5.4 \pm 0.5$ \\

        $10^3b_{\eta,\text{SiII(1260)}}$ & $-3.1 \pm 0.6$ & $-2.9 \pm 0.5$ & $-3.2 \pm 0.6$ & $-3.8 \pm 0.7$ & $-3.2 \pm 0.6$ \\
        
        $b_\text{HCD}$ & $-0.0574 \pm 0.0034$ & $-0.0491 \pm 0.0029$ & $-0.061 \pm 0.004$ & $-0.082 \pm 0.005$ & $-0.089 \pm 0.006$ \\
    
        $b_\text{q}$ & $3.97 \pm 0.10$ & $3.35 \pm 0.13$ & $3.99 \pm 0.13$ & $4.73 \pm 0.17$ & $4.35 \pm 0.23$ \\
        
        \drp ($h^{-1}$Mpc) & $-0.08 \pm 0.04$ & $-0.01 \pm 0.06$ & $-0.16 \pm 0.07$ & $-0.09 \pm 0.08$ & $-0.11 \pm 0.13$ \\
        
        $\sigma_\text{v}$ ($h^{-1}$Mpc) & $5.21 \pm 0.17$ & $4.79 \pm 0.25$ & $5.35 \pm 0.25$ & $5.45 \pm 0.35$ & $3.46 \pm 0.35$ \\
        
        $\xi_0^\text{TP}$ & $0.074 \pm 0.008$ & $0.066 \pm 0.013$ & $0.061 \pm 0.014$ & $0.114 \pm 0.023$ & $0.52 \pm 0.09$ \\ 

        $10^4A_\text{inst}$ & $1.33 \pm 0.09$ & $1.4 \pm 0.1$ & $1.36 \pm 0.10$ & $1.35 \pm 0.10$ & $1.51 \pm 0.10$ \\
        
        \hline 

        $\chi^2$ & 11941.2 & 11920.1 & 11883.1 & 11624.9 & 11573.8 \\
        
        $\chi^2$/dof & 1.027 & 1.025 & 1.022 & 0.999 & 0.995 \\
        
        $z_\text{eff}$ & 2.32 & 2.09 & 2.33 & 2.61 & 2.98 \\
         
    \end{tabular}
    \end{adjustbox}
    \caption{Results of the fits using the updated quasar templates for the new baseline and four redshift bins. The number of quasars in each bin is given, along with the redshift range of each bin where applicable. The free parameters are the same as described in section \ref{sec:results}. In all fits the BAO parameters are fixed at 1.0. Each fit has 11,644 data bins, 13 free parameters and we also list the effective redshift, $z_\text{eff}$. Since each fit is a joint fit of the auto- and cross-correlations, we list the $\chi^2$ and reduced $\chi^2$ ($\chi^2$/dof) values given for the joint fit.}
    \label{table:iron}
\end{table}

We find \drp = $-0.08 \pm 0.04$\mpc and $\sigma_\text{v} = 5.21 \pm 0.17$\mpc, which are shown as the red shaded region in figure \ref{fig:zevol_iron}. When testing for any evolution on \drp with redshift we find that the bias on \drp has been resolved. For the four redshift bins we find \drp~[$h^{-1}$Mpc]~=~$-0.01 \pm 0.06$, $-0.16 \pm 0.07$, $-0.09 \pm 0.08$, and $-0.11 \pm 0.13$, respectively. The results for $\sigma_\text{v}$ also suggest a better precision on the measured redshifts. We find $\sigma_\text{v}$ [$h^{-1}$Mpc] = $4.79 \pm 0.25$, $5.35 \pm 0.25$, $5.45 \pm 0.35$, and $3.46 \pm 0.35$ for the four redshift bins, respectively. 

Our updated \texttt{HIZ} templates and Redrock code have corrected the bias that was found using previous versions of the quasar templates and Redrock. The updated templates and Redrock will provide the redshifts for the DESI year one \lya cosmology studies \cite{desikp6}. The redshifts will be publicly released with the year one quasar sample.

\section{Summary}
\label{sec:conclusions}
We present here the study of the impact of quasar redshift errors on the 3D \lya forest-quasar cross-correlation. Using EDR+M2 data, the \lya forest sample from \cite{cesar}, and the auto-correlation from \cite{calum2023}, we compute the cross-correlation using a 1\mpc bin width up to 80\mpc in the \rp and \rt directions. We validate our choice to use a 1\mpc bin width with mock data. Our model has 13 free parameters, of which \drp and $\sigma_\text{v}$ are used to quantify any systematic quasar redshift errors.

In our baseline fit we find \drp = $-1.94 \pm 0.15$\mpc ($-205 \pm 15~\text{km~s}^{-1}$) and $\sigma_\text{v} = 5.8 \pm 0.5$\mpc ($610 \pm 50~\text{km~s}^{-1}$), which indicates that the EDR+M2 quasar redshifts are underestimated. It was reported in \cite{brodzeller2023} that there is a redshift-dependent bias in the eBOSS quasar templates that likely stems from the improper correction of the \lya forest optical depth. This introduces suppression in the \lya forest that increases with redshift. We confirm that this bias is present in the EDR+M2 sample by measuring and fitting the cross-correlation when the tracer quasar catalog is cut into four redshift bins with boundaries at $z=$ 2.2, 2.5, and 2.8. We find \drp [$h^{-1}$ Mpc] = $-1.07 \pm 0.17$, $-1.99 \pm 0.13$, $-3.5 \pm 0.4$, and $-3.4 \pm 1.0$ for the four bins, respectively. This shows a redshift-dependent bias that increases with redshift. 

New \texttt{HIZ} quasar templates were derived that corrected for the suppression from the \lya effective optical depth. A modified version of Redrock was run with these new templates to update the DESI Year 1 quasar sample at $z > 1.6$. We then repeat our analysis with the new Year 1 quasar redshifts to confirm that the bias in \drp is resolved. We find \drp~=~$-0.08 \pm 0.04$~\mpc ($-9.0 \pm 4.0~\text{km~s}^{-1}$) and $\sigma_\text{v} = 5.21 \pm 0.17$\mpc ($-548 \pm 17~\text{km~s}^{-1}$) when repeating the analysis on the entire year 1 quasar sample. We again split the catalog into four redshift bins with boundaries at $z=$~2.2,~2.5,~and 2.8 and we find \drp [$h^{-1}$ Mpc] = $-0.01 \pm 0.06$, $-0.16 \pm 0.07$, $-0.09 \pm 0.08$, and $-0.11 \pm 0.13$ for the four redshift bins, respectively. 

We have shown that our updated \texttt{HIZ} quasar templates have mitigated the redshift bias on \drp. These templates will be used to provide redshifts for the DESI year 1 quasar sample and those redshifts will be used for the Year 1 analyses.

\appendix
\section{Fit Parameters}
\label{sec:appfitpars}
\begin{table}[H]
    \begin{adjustbox}{width=\columnwidth}
    \centering
    \begin{tabular}{ccl}
    
    Parameter & Value & Description \\ \hline \hline
    $\alpha_\parallel, \alpha_\perp$ & 1.0 & BAO peak position parameters \\

    $A_{\text{peak}}$ & 1.0 & BAO amplitude (eq.~\ref{eq:PQL}) \\
    
    $b_\text{q}$ & floating & Quasar bias parameter (eq.~\ref{eq:bq_zdep}) \\
    
    $\beta_\text{q}$ & 0.323 & Quasar RSD parameter (derived) (eq.~\ref{eq:rsdqso}) \\

    $\alpha_\text{q}$ & 1.44 &  Redshift evolution parameter for $b_\text{q}$ (eq.~\ref{eq:bq_zdep}) \\
    
    $b_\text{Lya}$ & floating & Lya absorber bias parameter (eq.~\ref{eq:blya}) \\
    
    $\beta_\text{Lya}$ & floating & Lya absorber RSD parameter (\ref{eq:effbias})  \\

    $\alpha_\text{Lya}$ & 2.9 & Redshift evolution parameter for 
    $b_\text{Lya}$ (eq.~\ref{eq:bq_zdep}) \\
    
    $b_\text{HCD}$ & floating & bias parameter from unidentified HCD systems (eq.~\ref{eq:effbias}) \\
    
    $\beta_\text{HCD}$ & floating & RSD parameter from  unidentified HCD systems (eq.~\ref{eq:effbias}) \\
    
    $L_\text{HCD}$ & \mpch{10} & Length scale from unidentified HCD systems (eq.~\ref{eq:effbias}) \\
    
    $b_\text{m}$ & floating & bias parameters for the metal species listed in table~\ref{table:metals} \\
    
    $\beta_\text{m}$ & 0.5 & RSD parameters for the metal species listed in table~\ref{table:metals} \\

    $\alpha_\text{m}$ & 1.0 &  Redshift evolution parameter for metal biases $b_\text{m}$ \\
    
    $\sigma_v$ & floating & quasar non-linear velocity and precision (eq.~\ref{eq:FNL}) \\
    
    \drp & floating & the systematic offset in quasar redshift (eq.~\ref{eq:drp}) \\
    
    $f$ & 0.9703 & linear growth rate of structure (eq.~\ref{eq:sigma_f}) \\
    
    \textbf{$\Sigma$} = ($\Sigma_\parallel$, $\Sigma_\perp$) & (6.37, 3.26)\mpc & Non-linear broadening of the BAO peak (eq.~\ref{eq:PQL})  \\
    
    \textbf{R} = ($R_\parallel$, $R_\perp$) & (1, 1)\mpc & Smoothing parameters for the binning of the correlation (eq.~\ref{eq:Gk})  \\
    
    $\xi^\text{TP}_0$ & floating & Transverse proximity effect (eq.~\ref{eq:TPE}) \\
    
    $a_\text{UV}, t_\text{UV}$ & 0 & Quasar radiation and anisotropy lifetime (eq.~\ref{eq:TPE})  \\
    
    $\lambda_\text{UV}$ & \mpch{300} & mean free path of UV photons (eq.~\ref{eq:TPE}) \\

    $A_\text{inst}$ & floating & Instrumental systematics parameter (equation 4.11 in \cite{calum2023})

    \end{tabular}
    \end{adjustbox}
    \caption{Parameters of the joint fit of the \lya forest auto- and cross-correlations. The second column gives the value for each fixed parameter or if it is floating in the fit. The third column provides a brief description of each parameter and also a reference to the relevant equation. All biases $b$ refer to the bias calculated at the effective redshift ($z_\text{eff} = 2.34$ for the baseline fit).}
    \label{table:modelparams}
\end{table}

\acknowledgments
The work of Abby Bault and David Kirkby was supported by the U.S. Department of Energy, Office of Science, Office of High Energy Physics, under Award No. DE-SC0009920.

This material is based upon work supported by the U.S. Department of Energy (DOE), Office of Science, Office of High-Energy Physics, under Contract No. DE–AC02–05CH11231, and by the National Energy Research Scientific Computing Center, a DOE Office of Science User Facility under the same contract. Additional support for DESI was provided by the U.S. National Science Foundation (NSF), Division of Astronomical Sciences under Contract No. AST-0950945 to the NSF’s National Optical-Infrared Astronomy Research Laboratory; the Science and Technology Facilities Council of the United Kingdom; the Gordon and Betty Moore Foundation; the Heising-Simons Foundation; the French Alternative Energies and Atomic Energy Commission (CEA); the National Council of Science and Technology of Mexico (CONACYT); the Ministry of Science and Innovation of Spain (MICINN), and by the DESI Member Institutions: \url{https://www.desi.lbl.gov/collaborating-institutions}.~Any opi-nions, findings, and conclusions or recommendations expressed in this material are those of the author(s) and do not necessarily reflect the views of the U. S. National Science Foundation, the U. S. Department of Energy, or any of the listed funding agencies.

The authors are honored to be permitted to conduct scientific research on Iolkam Du’ag (Kitt Peak), a mountain with particular significance to the Tohono O’odham Nation.

\paragraph{Data Availability} 
All data points corresponding to the figures in this work, and the python code used to produce them, are available at \url{https://doi.org/10.5281/zenodo.10515085}.

%\begin{thebibliography}{99}
\bibliographystyle{JHEP.bst}
\bibliography{main}

%\end{thebibliography}

\end{document}